\documentstyle[11pt] {article}

\begin{document}

\title{Energy Transformations in a Relativistic Engine}

\author{Shailendra Rajput$^a$, Asher Yahalom$^{a,b}$ \& Hong Qin$^b$ \\
$^a$ Ariel University, Kiryat Hamada POB 3, Ariel 40700, Israel\\
$^b$ Princeton University, Princeton, New Jersey 08543, USA\\
e-mails: shailendrara@ariel.ac.il, asya@ariel.ac.il,\\ hongqin@princeton.edu}

\maketitle

\newcommand{\beq} {\begin{equation}}
\newcommand{\enq} {\end{equation}}
\newcommand{\ber} {\begin {eqnarray}}
\newcommand{\enr} {\end {eqnarray}}
\newcommand{\eq} {equation}
\newcommand{\eqs} {equations }
\newcommand{\mn}  {{\mu \nu}}
\newcommand{\sn}  {{\sigma \nu}}
\newcommand{\rhm}  {{\rho \mu}}
\newcommand{\sr}  {{\sigma \rho}}
\newcommand{\bh}  {{\bar h}}
\newcommand {\er}[1] {equation (\ref{#1}) }
\newcommand {\ern}[1] {equation (\ref{#1})}
\newcommand {\Ern}[1] {Equation (\ref{#1})}
\newcommand{\tR}  {{\tilde R}}
\newcommand{\atR}  {{|\tilde R|}}
\newcommand{\atRo}  {{|\tilde R_1|}}
\newcommand{\atRt}  {{|\tilde R_2|}}

\begin {abstract}
In a previous paper \cite{MTAY1} we have shown that Newt\-on'n third law cannot strictly
hold in a distributed system of which the different parts are at a finite distance from each other.
This is due to the finite speed of signal propagation which cannot exceed the speed of light at vacuum,
which in turn means that when summing the total force in the system the force does not add up to zero.
This was demonstrated in a specific example of two current loops with time dependent currents, the
above analysis led to suggestion of a relativistic engine \cite{MTAY3,AY1}.
Since the system is effected by a total force for a finite period of time this means that the system acquires mechanical momentum and energy, the question then arises how can we accomodate the law of momentum and energy conservation. The subject of momentum conversation was discussed in
\cite{MTAY4}, while preliminary results regarding energy conservation were discussed in \cite{AY2,RY,RY2}. Here we give a complete analysis of the exchange of energy between the mechanical part of the relativistic engine and the field part, the energy radiated from the relativistic engine is also discussed. We show that the relativistic engine effect on the energy is 4th-order in $\frac{1}{c}$ and no lower order relativistic engine effect on the energy exist.‎
\vspace*{0.5 cm}
\vspace*{0.5 cm}
\end {abstract}

\section{Introduction}

Special relativity is a theory of the structure of space-time.
It was introduced in Einstein's famous 1905 paper: "On the Electrodynamics of Moving Bodies" \cite{Einstein}.
This theory was a consequence of empiric observations and the laws of electromagnetism which were formulated in
the middle of the nineteenth century by Maxwell in his famous four partial
 differential equations \cite{Maxwell,Jackson,Feynman} which owe their current form to Oliver Heaviside \cite{Heaviside}.
One of the consequences of these equations is that an electromagnetic signal travels at the speed of light $c$,
which led people to believe that light is an electromagnetic wave. This was later used by Albert Einstein \cite{Einstein,Jackson,Feynman} to formulate his special theory of relativity which postulates that the speed of light in vacuum $c$ is the maximal allowed velocity in nature. According to the theory of relativity no object, message, signal (even if not electromagnetic) or field can travel faster than the speed of light in vacuum. Hence retardation, if someone at a distance $R$ from me changes something I may not know about it for at least a retardation time of $\frac{R}{c}$. This means that action and its reaction cannot be generated at the same time because of the signal finite propagation speed.

Newton's laws of motion are three physical laws that, together, laid the foundation for classical mechanics. They describe the relationship between a body and the forces acting upon it, and its motion in response to those forces. The three laws of motion were first compiled by Isaac Newton in his Philosophiae Naturalis Principia Mathematica (Mathematical Principles of Natural Philosophy), first published in 1687 \cite{Newton,Goldstein}. We will only be interested in this paper in the third law which states: When one body exerts a force on a second body, the second body \textbf{simultaneously} exerts a force equal in magnitude and opposite in direction on the first body.

According to the third law the total force in a system not affected by external forces is thus zero.
This law has numerous experimental verifications and seem to be one of the corner stones of physics. However, in light
of the previous discussion it is obvious that  action and its reaction cannot be generated at
the same time because of the finite speed of signal propagation, hence the
third law is false in an exact sense although it can be true for most practical application due to the high speed of signal propagation.
 Thus the total force cannot be null at a given time.

The locomotive systems of today are based on two material parts each obtaining momentum which is equal and opposite to the momentum
gained by the second part. A typical example of this type  of system is a rocket which sheds exhaust gas to propel itself. However, the above relativistic considerations suggest's a new type of motor in which the system is not composed of two material bodies but of a material body and field. Ignoring the field a naive observer will see the material body gaining
momentum created out of nothing, however, a knowledgeable observer will understand that the opposite amount of momentum is obtained
by the field as was shown in \cite{MTAY4}. Indeed Noether's theorem dictates that any system possessing
 translational symmetry will conserve momentum and the total physical system containing matter and field is indeed symmetrical under translations, while every sub-system (either matter or field) is not. This was already noticed by Feynman \cite{Feynman}. Feynman describes two orthogonally moving charges, apparently violating Newton's third law as the forces that the charges induce on each other do not cancel (last part of 26-2), this paradox is resolved in (27-6) in which it is shown that the momentum gained by the two charge system is balanced by the field momentum.

 In what follows we will assume that the magnetization and polarization of the medium are small and
therefore we neglect corrections to the Lorentz force suggested in \cite{Mansuripur}.

In a paper by Griffiths \&  Heald \cite{Griffiths} it was pointed out that strictly Coulomb’s law and the Biot-Savart law determine
the electric and magnetic fields for static sources only. Time-dependent generalizations of these two laws introduced by Jefimenko
\cite{Jefimenko} were used  to explore the applicability of Coulomb and Biot-Savart outside the static domain.

In a previous paper we used Jefimenko's \cite{Jefimenko,Jackson} equation to discuss the force between two current carrying coils \cite{MTAY1}.
This was later expanded to include the interaction between a current carrying loop and a permanent magnet  \cite{MTAY3,AY1}.
Since the system is affected by a total force for a finite period of time this means that the system acquires mechanical momentum and energy, the question then arises if we need to abandon the law of momentum and energy conservation. The subject of momentum conversation was discussed in
\cite{MTAY4}. In \cite{AY2,RY,RY2} some preliminary aspects of the exchange of energy between the mechanical part of the relativistic engine and the electromagnetic field were discussed. In particular it was shown that the electric energy expenditure is twice the kinetic energy
gained by the relativistic motor. It was also shown how some energy may be radiated from the relativistic engine device if the coils are not configure properly. In this paper we develop a methodology to deal with all aspects of the energy transformation in a relativistic engine.

The plan this paper is as follows: First we introduce the conservation of energy and momentum in a general electromagnetic system. Then we discuss the particular case of a simple relativistic engine made of two current loops, in which we shall consider the mechanical momentum and energy gained by engine. This will be followed by a general analysis of the energy transformations between the field and mechanical components in a relativistic engine. Which will be followed by a specific analysis of the two loop relativistic engine taking into account the various energy contributions expanded in powers of $\frac{1}{c}$,  in which $c$ is the speed of light in vacuum. We show that the relativistic engine effect on the energy is 4th-order in $\frac{1}{c}$ and no lower order relativistic engine effect on the energy exist.

\section{Energy \& Momentum Conservation}

Any system with space-time translational symmetry must conserve momentum and energy according to Noether's theorem. In the case of a system with
charge and current densities the energy-momentum conservation law takes the form \cite{Jackson}:
\beq
\int (\partial_\alpha \Theta^{\alpha \beta}+f^\beta) d^3 x = 0
\label{stencon}
\enq
$x^\alpha$ are space-time coordinates such that $\alpha,\beta \in \{0,1,2,3\}$,  $\partial_\alpha$ is a partial derivative with respect to the four dimensional coordinates and Einstein's summation convention is assumed, $d^3 x$ is a volume element.
In the above:
\beq
\Theta^{\alpha \beta} = \left(
                                \begin{array}{cc}
                                  e_{field} & \frac{1}{c}\vec S_p \\
                                  \frac{1}{c}\vec S_p & -T_{i j} \\
                                \end{array}
                              \right)
\label{sten}
\enq
and:
\beq
\int f^\beta d^3 x =(\frac{d E_{mech}}{dt},\frac{d \vec P_{mech}}{dt}).
\label{f}
\enq
The various terms in the matrix appearing in \ern{sten} are defined in terms of the electric field $\vec E$ and magnetic flux density
$\vec B$ as follows. The field energy density $e_{field}$ is defined such that:
\beq
E_{field}\equiv \int e_{field} d^3 x =\frac{\epsilon_0}{2} \int \left( \vec E^2 + c^2 \vec B^2 \right) d^3 x
\label{Efield}
\enq
in the above $\epsilon_0$ is the vacuum permittivity ($\simeq 8.85 \ 10^{-12} \ {\rm F \ m^{-1}}$).
Poynting's vector is defined as:
\beq
\vec S_p= \frac{1}{\mu_0} \vec E \times \vec B
\label{Poynting}
\enq
$\mu_0 =4 \pi \ 10^{-7}$ is the vacuum magnetic permeability. $T_{i j}$ is the Maxwell stress tensor:
\beq
T_{i j} = \epsilon_0 \left[E_i E_j + c^2  B_i B_j -
 \frac{1}{2} (\vec E^2 + c^2 \vec B^2) \delta_{i j} \right]
\label{Maxstr}
\enq
$i,j\in \{1,2,3\}$ and $\delta_{i j}$ is Kronecker's delta. \Ern{f} contains the mechanical energy and momentum
$E_{mech},\vec P_{mech}$ and the temporal derivative $\frac{d}{dt}$. Next we shall write the matrix \ern{stencon} in terms
of the spatial and temporal components separately.
The spatial components will yield the equation:
\beq
\frac{d P_{mech~i}}{dt}+\frac{d  P_{field~i}}{dt}= \oint_S T_{i j} {\hat n}_{j} da
\label{Pcon}
\enq
In the above $P_{field~i}$ is the $i$ component of the field momentum of the system:
\beq
\vec P_{field}=\epsilon_0 \int \vec E \times \vec B d^3 x
\label{Pfield}
\enq
$S$ is a closed surface encapsulating the volume in which the system is located, $\hat n$ is a unit vector normal to the surface, $da$ is a surface element.

\Ern{Pcon} as proved in \cite{Jackson} is a precise statement of momentum conservation in
a relativistic engine and from a pure point of view nothing else is needed, however, for the sake of concreteness a specific
example for the two current loops relativistic engine was analyses in  \cite{MTAY4} and will not be repeated here.

To conclude this section we shall look at the zeroth component of \ern{stencon}, which yields the energy conservation equation:
\beq
\frac{d E_{mech}}{dt}+\frac{d  E_{field}}{dt}= -\oint_S \vec S_p \cdot \hat n da.
\label{Econ}
\enq
The derivative of the mechanical energy is the power needed to sustain the system and is given by \cite{Jackson}:
\beq
Power \equiv  \frac{d E_{mech}}{dt}= \int d^3 x \vec J \cdot \vec E
\label{Econ2a}
\enq
$\vec J$ is the current flux density.  Hence:
\beq
\int d^3 x \vec J \cdot \vec E + \frac{d  E_{field}}{dt}= -\oint_S \vec S_p \cdot \hat n da.
\label{Econ2b}
\enq
Or:
\beq
\int d^3 x \vec J \cdot \vec E = - \frac{d  E_{field}}{dt} -\oint_S \vec S_p \cdot \hat n da.
\label{Econ2c}
\enq
In the following we will discuss the manifestation of energy conservation  as described by \ern{Econ2c} for a relativistic engine.

\section{The case of two current loops }
\label{twocurlo}
\begin{figure}
\vspace{3cm} \includegraphics{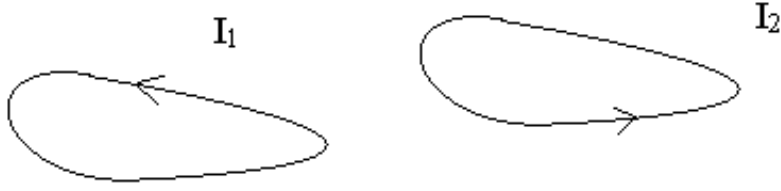}
\caption {Two current loops.}
 \label{twoloops}
\end{figure}

Consider two wires having segments of length $d\vec l_1,d\vec l_2$ located at $\vec x_1,\vec x_2$  respectively and
carrying currents $I_1, I_2$ (see figure \ref{twoloops}).
According to \cite{MTAY1} (equation (38)) the force on loop $_2$ generated by loop $_1$ takes the form:
\beq
\vec F_{21} = \frac{\mu_0}{4 \pi} I_2 (t) \sum_{n=0}^{\infty} \frac{I_1^{(n)}(t)}{n!}
(- \frac{1}{c})^n (1-n) \oint \oint R_{12}^{n-3} \vec R_{12} (d\vec l_2 \cdot d\vec l_1)
\label{F21ft3}
\enq
in which $\vec R_{12} \equiv \vec x_1 - \vec x_2$, $R_{12} \equiv |\vec R_{12}|$. As in all expansions the
above equation is valid only for a certain environment of $t$ on the time axis which depends on the function $I_1 (t)$, this enviroment shall be defined using the convergence radius $T_{max}$. That is \ern{F21ft3} is valid only in the domain $[t-T_{max},t+T_{max}]$.
We note that there is no first order contribution to the force. Hence the next contribution to
the force after the quasi-static term is second order.
Let us define the dimensionless geometrical factor $\vec K_{21n}$ as:
\beq
\vec K_{21n} = \frac{1}{h^n}  \oint \oint R_{12}^{n-3} \vec R_{12} (d\vec l_2 \cdot d\vec l_1) = -\vec K_{12n}.
\label{Kdef}
\enq
in the above $h$ is some characteristic distance between the coils. In terms of $\vec K_{21n}$ we can write \ern{F21ft3} as:
\beq
\vec F_{21} = \frac{\mu_0}{4 \pi} I_2 (t) \sum_{n=0}^{\infty} \frac{I_1^{(n)}(t)}{n!} (- \frac{h}{c})^n (1-n) \vec K_{21n}.
\label{F21ft4}
\enq
The force due to coil $2$ that acts on coil $1$ is:
\beq
\vec F_{12} = \frac{\mu_0}{4 \pi} I_1 (t) \sum_{n=0}^{\infty} \frac{I_2^{(n)}(t)}{n!} (- \frac{h}{c})^n (1-n) \vec K_{12n}.
\label{F12ft}
\enq
The total force on the system is thus:
\ber
&&\hspace{-1 cm} \vec F_{T} = \vec F_{12} + \vec F_{21} =
\nonumber \\
&&\hspace{-1 cm}\frac{\mu_0}{4 \pi}  \sum_{n=0}^{\infty} \frac{(1-n)}{n!} (- \frac{h}{c})^n  \vec K_{12n}
\left(I_1 (t) I_2^{(n)}(t) - I_2 (t) I_1^{(n)}(t)\right) .
\label{Ftotal2}
\enr
We note that the quasi-static term $n=0$ does not contribute to the sum nor does the $n=1$ term.
The fact that the retarded field "corrects" itself to first order in order to "mimic" a non retarded
field was already noticed by Feynman \cite{Feynman}.
Hence we can write:
\beq
\vec F_{T} = \frac{\mu_0}{4 \pi}  \sum_{n=2}^{\infty} \frac{(1-n)}{n!} (- \frac{h}{c})^n  \vec K_{12n}
 \left(I_1 (t) I_2^{(n)}(t) - I_2 (t) I_1^{(n)}(t)\right) .
\label{Ftotal3}
\enq
We conclude that in general Newton's third law is not satisfied, taking the leading non-vanishing terms in the above sum
we obtain:
\ber
\hspace{-0.5 cm}\vec F_{T} &\cong& -\frac{\mu_0}{8 \pi} (\frac{h}{c})^2  \vec K_{122}
\left(I_1 (t) I_2^{(2)}(t) - I_2 (t) I_1^{(2)}(t)\right) .
\label{Ftotal4}
\enr
This result is correct up to a second order in $\frac{1}{c}$. Assuming that the total momentum of the system and the current derivatives are null at $t=0$, we obtain a mechanical linear momentum as follows:
\ber
&&\vec P_{mech}=\int_0^t \vec F_{T} (t')dt' \cong
\nonumber \\
&& -\frac{\mu_0}{8 \pi} (\frac{h}{c})^2  \vec K_{122}
\left(I_1 (t) I_2^{(1)}(t) - I_2 (t) I_1^{(1)}(t)\right).
\label{Pmech1}
\enr
For simplicity we will now on assume a permanent current in loop $_2$ hence:
\beq
\vec P_{mech} \cong \frac{\mu_0}{8 \pi} I_1^{(1)}(t) I_2 (\frac{h}{c})^2  \vec K_{122} \propto \frac{1}{c^2}.
\label{Pmech1b}
\enq
For a calculation of $\vec K_{122}$ in particular geometries see \cite{MTAY1,MTAY3,AY1}. We stress  that
the mechanical momentum  given is of order $\frac{1}{c^2}$ and higher order terms are neglected.
 The kinetic mechanical energy associated with this momentum is:
\beq
E_{mech}= \frac{ \vec P_{mech}^2}{2 M} = \frac{1}{2} \vec P_{mech} \cdot \vec v \propto \frac{1}{c^4}.
\label{Emech}
\enq
Were $M$ is the mass of the relativistic engine and:
\beq
\vec v =  \frac{\vec P_{mech}}{M} \propto \frac{1}{c^2}
\label{vecv}
\enq
is the engine velocity. This indicates that unlike the conservation of momentum \cite{MTAY4} which was independent of the mass and therefore of the velocity attained by the system, in the calculations of both the mechanical and electromagnetic energies the systems velocity and mass are of paramount importance. We also note that the expression for mechanical energy is of order $\frac{1}{c^4}$, lower order corrections do not exist and higher order corrections are neglected.

We define the relativistic engine effect as a motion of the ‎center of mass of a system due to the interaction of its two ‎subsystems, if there is no motion of the center of mass than ‎the effect is null. It may be that the center of mass is moving ‎due to the motion of each subsystem separately but then, ‎according to the above definition this is not a relativistic engine ‎effect but something else.

\section{Field Energy}

Consider two sub-systems denoted system $_1$ and system $_2$ which are far apart such that their interaction is negligible.
In this case \er{Econ2c} is correct for each sub-system separately, that is:
\beq
\int d^3 x \vec J_1 \cdot \vec E_1 = - \frac{d  E_{field~1}}{dt} -\oint_S \vec S_{p~1} \cdot \hat n da.
\label{Econ1}
\enq
\beq
\int d^3 x \vec J_2 \cdot \vec E_2 = -\frac{d  E_{field~2}}{dt} -\oint_S \vec S_{p~2} \cdot \hat n da.
\label{Econ2}
\enq
Next we will put the two loops closer together such that they may interact but without modifying the
charge and the current densities of each of the subsystems. The total fields of the combined system are:
\beq
\vec E = \vec E_1 + \vec E_2, \qquad \vec B = \vec B_1 + \vec B_2
\label{Fields}
\enq
Since both the field energy \ern{Efield} and Poynting's vector \ern{Poynting}
are quadratic in the fields the following result is obtained:
\ber
E_{field} &\equiv& \frac{\epsilon_0}{2} \int \left( \vec E^2 + c^2 \vec B^2 \right) d^3 x
=E_{field~1}+E_{field~2}+E_{field~12}
\nonumber \\
E_{field~1} &\equiv& E_{E field~1}+E_{M field~1}
\equiv \frac{\epsilon_0}{2} \int \left( \vec E_1^2 + c^2 \vec B_1^2 \right) d^3 x
\nonumber \\
E_{field~2} &\equiv& E_{E field~2}+E_{M field~2}
\equiv \frac{\epsilon_0}{2} \int \left( \vec E_2^2 + c^2 \vec B_2^2 \right) d^3 x
\nonumber \\
E_{field~12} &\equiv& E_{E field~12}+E_{M field~12}
\nonumber \\
&\equiv& \epsilon_0 \int \left(  \vec E_1 \cdot \vec E_2 + c^2 \vec B_1 \cdot \vec B_2 \right) d^3 x
\label{Efielddiv}
\enr
\ber
\vec S_p & \equiv & \frac{1}{\mu_0} \vec E \times \vec B  = \vec S_{p~1}+\vec S_{p~2}+ \vec S_{p~12}
 \nonumber \\
\vec S_{p~1} & \equiv & \frac{1}{\mu_0} \vec E_1 \times \vec B_1
 \nonumber \\
\vec S_{p~2} &\equiv & \frac{1}{\mu_0} \vec E_2 \times \vec B_2
 \nonumber \\
\vec S_{p~12} & \equiv & \frac{1}{\mu_0} \left( \vec E_1 \times \vec B_2 +\vec E_2 \times \vec B_1\right)
\label{Poyntingdiv}
\enr
The power invested in the combined system in bilinear in the current flux density and the electric field according to \ern{Econ2a}.
This will lead to the following expression:
\ber
Power &=& \int d^3 x \vec J \cdot \vec E = Power_1 + Power_2 + Power_{12}
 \nonumber \\
Power_1 & \equiv & \int d^3 x \vec J_1 \cdot \vec E_1
 \nonumber \\
Power_2 & \equiv & \int d^3 x \vec J_2 \cdot \vec E_2
 \nonumber \\
Power_{12} &\equiv & \int d^3 x \left(\vec J_1 \cdot \vec E_2 + \vec J_2 \cdot \vec E_1 \right)
 \label{Powerpart}
\enr
Subtracting from \ern{Econ2c} the expressions given in \ern{Econ1} and \ern{Econ2}:
\ber
& &\hspace{-1cm}Power - Power_1 - Power_2
 \nonumber \\
&\hspace{-1cm}=&\hspace{-1cm} -\frac{d  (E_{field}-E_{field~1}-E_{field~2})}{dt}  -\oint_S \left(\vec S_{p}-\vec S_{p~1}-\vec S_{p~2} \right) \cdot \hat n da.
\label{Econt-1}
\enr
taking into account \ern{Efielddiv}, \ern{Poyntingdiv} and \ern{Powerpart} we arrive at:
\beq
Power_{12} = -\frac{d  E_{field~12}}{dt} -\oint_S \vec S_{p~12} \cdot \hat n da.
\label{Econt0}
\enq

\section{Field Energy of a System of two current loops}

We now return to the case of two current loops as described in section \ref{twocurlo}. However, now
we assume two types of time dependence. One is due to the intrinsic time dependence of the current flowing
through the loop and another is due to its movement as part of the relativistic engine. Thus the current density can be
written as:
\beq
\vec J(\vec x ,t) = \vec J' (\vec x - \vec x_c (t) ,t),
\label{Jp}
\enq
in which $\vec J' (\vec x ,t)$ is the current density in the moving frame of the relativistic engine and
$\frac{d \vec x_c (t)}{dt} = \vec v (t) $ is the velocity of that frame which is also the velocity of the relativistic engine.
The vector potential is given as \cite{Jackson}:
\beq
\vec A (\vec x,t) = \frac{\mu_0}{4 \pi} \int d^3 x' \frac{\vec J (\vec x',t_{ret})}{R}, \ \vec R \equiv \vec x-\vec x',
\ t_{ret} \equiv t-\frac{R}{c}.
\label{A}
\enq
This can be written as a power series in terms of $-\frac{R}{c}$ in the form:
\ber
\vec A (\vec x,t) &=& \frac{\mu_0}{4 \pi} \sum_{n=0}^{\infty} \frac{1}{n!} \int d^3 x' \frac{1}{R} (- \frac{R}{c})^n  \frac{d^n}{dt^n} \vec J (\vec x',t)
\nonumber \\
\hspace{-1cm} &\hspace{-3cm}=& \hspace{-1.5cm} \frac{\mu_0}{4 \pi} \sum_{n=0}^{\infty} \frac{1}{n!} \frac{d^n}{dt^n}  \int d^3 x' \frac{1}{R} (- \frac{R}{c})^n \vec J (\vec x',t)
\nonumber \\
\hspace{-1cm} &\hspace{-3cm} =&  \hspace{-1.5cm}\frac{\mu_0}{4 \pi} \sum_{n=0}^{\infty} \frac{1}{n!} \frac{d^n}{dt^n}  \int d^3 x' \frac{1}{R} (- \frac{R}{c})^n  \vec J' (\vec x' - \vec x_c (t) ,t),
\label{A2}
\enr
As in the case of \ern{F21ft3} this expansion is only valid in the domain $[t-T_{max},t+T_{max}]$ in which $T_{max}$ being the "convergence radius" of the above series. Hence this is only valid for distances satisfying $\frac{R}{c}<T_{max}$ or:
\beq
R < c \ T_{max} \equiv R_{max}
\label{Rmax}
\enq
since $c$ is a big number we will occasionally use \ern{A2} for infinite distances while remembering that although $R_{max}$ may be big it is not infinite.
Let us introduce a comoving integration variable: $\tilde{\vec{x}}=\vec x' - \vec x_c (t), \vec x'=\tilde{\vec{x}}+\vec x_c (t)$
such that: $R(t)=|\vec x' - \vec x|=|\tilde{\vec{x}}+\vec x_c (t)-\vec x|$ in such a comoving coordinate system we have:
\beq
\vec A (\vec x,t) = \frac{\mu_0}{4 \pi} \sum_{n=0}^{\infty} \frac{1}{n!} \frac{d^n}{dt^n}  \int d^3 \tilde{x}\frac{1}{R(t)} (- \frac{R(t)}{c})^n  \vec J' (\tilde{\vec{x}} ,t),
\label{A3}
\enq
For a thin and uniform current loop this can be written as:
\beq
\vec A (\vec x,t) = \frac{\mu_0}{4 \pi} \sum_{n=0}^{\infty} \frac{1}{n!} \frac{d^n}{dt^n} \left[ I (t) \oint d \tilde{\vec{l}} \frac{1}{R(t)} (- \frac{R(t)}{c})^n \right].
\label{A4}
\enq
Let us define:
\ber
\vec A^{(n)} (\vec x,t) &=& \frac{\mu_0}{4 \pi n!} \frac{d^n}{dt^n} \left[ I (t) \oint d \tilde{\vec{l}} \frac{1}{R(t)} (- \frac{R(t)}{c})^n \right]
\nonumber \\
&=& (-1)^n \frac{\mu_0}{4 \pi n! c^n} \frac{d^n}{dt^n} \left[ I (t) \oint d \tilde{\vec{l}} R(t)^{n-1} \right] ,
\label{A4b}
\enr
in terms of $\vec A^{(n)}$ we may write:
\beq
\vec A (\vec x,t) = \sum_{n=0}^{\infty} \vec A^{(n)} (\vec x,t).
\label{A4c}
\enq
We notice $\vec A^{(n)}$ is at least of order $\frac{1}{c^n}$ but may contain higher order terms. The reason for this is the fact that
$\vec A^{(n)}$ contains temporal derivatives of $R(t)$ which satisfy the equation:
\beq
\frac{d R (t)}{dt} = - \frac{\vec R (t)}{R (t)} \cdot \vec v = -\hat R \cdot \vec v, \qquad \hat R \equiv \frac{\vec R (t)}{R (t)}
\label{dR0}
\enq
and $\vec v$  is proportional to $\frac{1}{c^2}$ for a relativistic engine according to \ern{vecv}. Thus $\vec A^{(n)}$ may contain terms which
are up to order of $(\frac{1}{c})^{3n-2}$. We also notice that the zeroth order terms takes the form:
\beq
\vec A^{(0)} (\vec x,t) = \frac{\mu_0}{4 \pi} I (t)  \oint d \tilde{\vec{l}} \frac{1}{R(t)},
\label{A5}
\enq
which is just the quasi static approximation to $\vec A$. It also easy to see that the first order term:
\beq
\vec A^{(1)} (\vec x,t) = 0
\label{A1null}
\enq
 as $\oint d \tilde{\vec{l}} = 0$. As $\vec A^{(n)}$ contains terms in various orders of $\frac{1}{c}$, and we will be interested in analyzing the energy problem for a definite order of $\frac{1}{c}$, we introduce the square bracket notation $G^{[n]}$ to denote a quantity of order $(\frac{1}{c})^{n}$. Obviously, generally  $\vec A^{[n]} \neq \vec A^{(n)}$.
The electric field $\vec E$ can be calculated (in the gauge $\Phi=0$)  as \cite{Jackson}:
\beq
\vec E = - \partial_t \vec A.
\label{E1}
\enq
We shall define for computational convenience:
\beq
\vec E^{(n)} \equiv - \partial_t \vec A^{(n)}.
\label{E1b}
\enq
 This means that $\vec E ^{[n]} \neq \vec E^{(n)}$ will contain terms coming from $\vec A^{[n]} $ but also from $\vec A^{[n-2]}$  as the derivative will create terms which have an additional $(\frac{1}{c})^{2}$ factor. As for the magnetic field $\vec B$ which can be calculated as \cite{Jackson}:
\beq
\vec B = \vec \nabla \times \vec A
\label{Bgeneral0}
\enq
it is easy to see that $\vec B^{[n]} $ is the same order as $\vec A^{[n]}$ and:
\beq
\vec B^{[n]} = \vec \nabla \times \vec A^{[n]}
\label{Bgeneraln}
\enq
In what follows we will use the above expressions to analyzed the energy transformation order by order starting from $n=0$ and up to
 $n=4$ which according to \ern{Emech} is the most relevant for analyzing the energy transfer between the field and the mechanical components of a relativistic engine.

 \subsection{Normalization}
  To avoid carrying the factor $\frac{\mu_0}{4 \pi}$ in numerous calculations we shall define:
 \beq
 \vec A' \equiv \vec A \frac{4 \pi}{\mu_0} \Rightarrow  \vec A = \vec A' \frac{\mu_0}{4 \pi}.
 \label{Ap}
  \enq
And hence:
\beq
 \vec E' \equiv \vec E \frac{4 \pi}{\mu_0} \Rightarrow  \vec E = \vec E' \frac{\mu_0}{4 \pi}, \qquad
  \vec B' \equiv \vec B \frac{4 \pi}{\mu_0} \Rightarrow  \vec B = \vec B' \frac{\mu_0}{4 \pi}
 \label{EBp}
  \enq
The field energy (see \ern{Efielddiv}) can be calculated in terms of the new variables as:
\beq
E_{field~12} = \frac{\mu_0}{16 \pi^2} \int \left( \frac{1}{c^2} \vec E'_1 \cdot \vec E'_2 +  \vec B'_1 \cdot \vec B'_2 \right) d^3 x
\label{Efielddivp1}
\enq
where we are reminded that $c^2 = \frac{1}{\mu_0 \epsilon_0}$, the Pointing vector (\ern{Poyntingdiv}) will take the following form in terms of the new variables:
\beq
\vec S_{p~12} = \frac{\mu_0}{16 \pi^2} \left( \vec E'_1 \times \vec B'_2 +\vec E'_2 \times \vec B'_1\right).
\label{Poyntingdivp}
\enq
Finally the mechanical work which is represented by \ern{Powerpart} will take the form:
\beq
Power_{12} = \frac{\mu_0}{4 \pi} \int d^3 x \left(\vec J_1 \cdot \vec E'_2 + \vec J_2 \cdot \vec E'_1 \right).
\label{Powerpartp}
\enq
Now looking at \ern{Econt0} the following normalization is suggested:
\ber
\vec S'_{p~12} &\equiv& \frac{16 \pi^2}{\mu_0} \vec S_{p~12}  \Rightarrow  \vec S_{p~12} = \frac{\mu_0}{16 \pi^2} \vec S'_{p~12}
\nonumber \\
\vec S'_{p~12} &=& \left( \vec E'_1 \times \vec B'_2 +\vec E'_2 \times \vec B'_1\right).
\label{Poyntingdivp2}
\enr
\ber
E'_{field~12} &\equiv& \frac{16 \pi^2}{\mu_0} E_{field~12}  \Rightarrow  E_{field~12} = \frac{\mu_0}{16 \pi^2} E'_{field~12}
\nonumber \\
E'_{field~12} &=& \int \left( \frac{1}{c^2} \vec E'_1 \cdot \vec E'_2 +  \vec B'_1 \cdot \vec B'_2 \right) d^3 x
\label{Efielddivp2}
\enr
\ber
Power'_{12} &\equiv& \frac{16 \pi^2}{\mu_0} Power_{12}  \Rightarrow  Power_{12} = \frac{\mu_0}{16 \pi^2} Power'_{12}
\nonumber \\
Power'_{12} &=& 4 \pi  \int d^3 x \left(\vec J_1 \cdot \vec E'_2 + \vec J_2 \cdot \vec E'_1 \right).
\label{Powerpartp2}
\enr
In terms of the normalized quantities we may write the relativistic engine energy transformation \ern{Econt0} as:
\beq
Power'_{12} = -\frac{d  E'_{field~12}}{dt} -\oint_S \vec S'_{p~12} \cdot \hat n da.
\label{Econt0p}
\enq
This can be analyzed order by order in terms of powers of $\frac{1}{c}$ hence:
\beq
Power'^{[n]}_{12} = -\frac{d  E'^{[n]}_{field~12}}{dt} -\oint_S \vec S'^{[n]}_{p~12} \cdot \hat n da.
\label{Econt0pn}
\enq
We notice that because of the perfector  $\frac{1}{c^2}$ of the electric field contribution in \ern{Efielddivp2} this term will not
contribute to the field energy for the lowest orders: $n=0,1$. We also notice that since we are considering a case of two loop currents
\ern{Powerpartp2} takes the form:
\beq
Power'_{12} = 4 \pi   \left( I_1 (t) \oint d \tilde{\vec{l_1}} \cdot \vec E'_2 (\vec x_1) + I_2 (t) \oint d \tilde{\vec{l_2}} \cdot \vec E'_1 (\vec x_2) \right).
\label{Powerpartp2lo}
\enq
We are now ready to analyzed the field contribution order by order.

\subsection{$n=0$}

Let us look at \ern{Econt0pn} and study it for the zeroth order in $\frac{1}{c}$:
\beq
Power'^{[0]}_{12} = -\frac{d  E'^{[0]}_{field~12}}{dt} -\oint_S \vec S'^{[0]}_{p~12} \cdot \hat n da.
\label{Econt0ord0}
\enq

\subsubsection{Power}

We shall start by calculating $Power_{12}^{[0]}$, according to \ern{Powerpartp2lo}:
\beq
Power'^{[0]}_{12} =  4 \pi   \left( I_1 (t) \oint d \tilde{\vec{l_1}} \cdot \vec E'^{[0]}_2 (\vec x_1) + I_2 (t) \oint d \tilde{\vec{l_2}} \cdot \vec E'^{[0]}_1 (\vec x_2) \right).
\label{pow0}
\enq
The field $\vec E'^{[0]}$ can only come from the vector potential $\vec A'^{[0]}=\vec A'^{(0)}$ as all other $\vec A'^{(n)}$ are of higher order in $n$, hence according to \ern{A5}:
\beq
\vec A'^{[0]}(\vec x,t)=\vec A'^{(0)} (\vec x,t) =  I (t)  \oint d \tilde{\vec{l}} \frac{1}{R(t)}.
\label{Ab0}
\enq
According to \ern{E1b}:
\beq
\vec E'^{(0)} = - \partial_t \vec A'^{(0)} = - \partial_t I (t)  \oint d \tilde{\vec{l}} \frac{1}{R(t)} -
I (t) \oint d \tilde{\vec{l}}\quad \frac{\vec v \cdot \vec R(t)}{R^3(t)}.
\label{E0b}
\enq
However, the second term in the above equation is of order of $\frac{1}{c^2}$ and will be dealt with later. Hence:
\beq
\vec E'^{[0]} = -  \partial_t I (t)  \oint d \tilde{\vec{l}} \frac{1}{R(t)}.
\label{E0}
\enq
we notice that the derivative of the integral is of order $(\frac{1}{c})^2$ hence does not contribute to the zeroth order. Now since the second coil has a constant current this leads to the result:
\beq
\vec E'^{[0]}_2 = 0.
\label{E02}
\enq
Inserting the above result into \ern{pow0} yields:
\beq
Power'^{[0]}_{12} =  4 \pi   \ I_2 (t) \oint d \tilde{\vec{l_2}} \cdot \vec E'^{[0]}_1 (\vec x_2).
\label{pow02}
\enq
Inserting \ern{E0} into \ern{pow02} will yield the expression:
\beq
Power'^{[0]}_{12}  =  -  4 \pi \partial_t I_1 (t) I_2 \oint \oint d \tilde{\vec{l}_1} \cdot d \tilde{\vec{l}_2} \frac{1}{R_{12}(t)},
\qquad \vec R_{12} = \vec x_1 - \vec x_2
\label{pow04}
\enq
This can be written in terms of the familiar mutual inductance \cite{Jackson}:
\beq
M^{[0]}_{12} \equiv \frac{\mu_0}{4 \pi} \oint d \vec l_1 \cdot \oint d \vec l_2 \frac{1}{|\vec x_1 - \vec x_2|}
\label{mutind}
\enq
as:
\beq
Power_{12}^{[0]} = \frac{\mu_0}{16 \pi^2} Power'^{[0]}_{12} = -  \partial_t I_1 (t) I_2  M^{[0]}_{12}
\label{pow05}
\enq
We notice that $Power_{12}^{[0]}$ may be positive or negative according to the relative position of the current loops and current directions, hence, energy can be invested or extracted from the combined system according to the system configuration.

\subsubsection{Field Energy}

Turning our attention next to field energy defined in \ern{Efielddivp2}:
\beq
E'^{[0]}_{field~12} = \int \vec B'^{[0]}_1 \cdot \vec B'^{[0]}_2  d^3 x.
\label{Efielddiv0}
\enq
$\vec B'^{[0]}$ is calculated according to \ern{Bgeneraln} as:
\beq
\vec B'^{[0]} = \vec \nabla \times \vec A'^{[0]}.
\label{Bgeneraln0}
\enq
Now since $ \vec A'^{[0]}=\vec A'^{(0)}$ we may use \ern{A5} to obtain:
\beq
\vec B'^{[0]} (\vec x,t) = -  I (t)  \oint d \tilde{\vec{l}} \times  \vec \nabla \frac{1}{R (t)}=
 I (t)  \oint d \tilde{\vec{l}} \times  \frac{\vec R(t) }{R^3 (t)}.
\label{B0}
\enq
Inserting \ern{B0} into \ern{Efielddiv0} will yield:
\beq
E'^{[0]}_{field~12} =  I_1 (t)  I_2   \int d^3 x  \oint d \tilde{\vec{l_1}} \times  \vec \nabla \frac{1}{R_1 (t)} \cdot \oint d \tilde{\vec{l_2}} \times  \vec \nabla \frac{1}{R_2 (t)}
\label{Efielddiv03}
\enq
in which we recall that $I_2$ is time independent. Using a well known identity from vector analysis we may write:
\ber
E'^{[0]}_{field~12} &=&   I_1 (t)  I_2   \int d^3 x  \oint \oint [
(d \tilde{\vec{l_1}} \cdot d \tilde{\vec{l_2}}) (\vec \nabla \frac{1}{R_1 (t)} \cdot  \vec \nabla \frac{1}{R_2 (t)})
\nonumber \\
 & - & (d \tilde{\vec{l_1}} \cdot \vec \nabla \frac{1}{R_2 (t)}) (d \tilde{\vec{l_2}} \cdot \vec \nabla \frac{1}{R_1 (t)}) ].
\label{Efielddiv04}
\enr
We shall show in the appendix A that:
\beq
\int d^3 x  \oint \oint [ (d \tilde{\vec{l_1}} \cdot \vec \nabla \frac{1}{R_2 (t)}) (d \tilde{\vec{l_2}} \cdot \vec \nabla \frac{1}{R_1 (t)}) ]
= 0
\label{Efielddiv05}
\enq
Hence:
\beq
E'^{[0]}_{field~12} =   I_1 (t)  I_2   \int d^3 x  \oint \oint [
(d \tilde{\vec{l_1}} \cdot d \tilde{\vec{l_2}}) (\vec \nabla \frac{1}{R_1 (t)} \cdot  \vec \nabla \frac{1}{R_2 (t)})].
\label{Efielddiv06}
\enq
Now:
\beq
\vec \nabla \frac{1}{R_1 (t)} \cdot  \vec \nabla \frac{1}{R_2 (t)} = \vec \nabla \cdot (\frac{1}{R_1 (t)} \vec \nabla \frac{1}{R_2 (t)} )
- \frac{1}{R_1 (t)} \vec \nabla^2 \frac{1}{R_2 (t)}
\label{Efielddiv07}
\enq
Taking into account that \cite{Jackson}:
\beq
\vec \nabla^2 \frac{1}{R_2 (t)} = - 4 \pi \delta (\vec R_2)
\label{Efielddiv08}
\enq
in which $\delta (\vec R_2)$ is a three dimensional delta function, we have:
\beq
\vec \nabla \frac{1}{R_1 (t)} \cdot  \vec \nabla \frac{1}{R_2 (t)} = \vec \nabla \cdot (\frac{1}{R_1 (t)} \vec \nabla \frac{1}{R_2 (t)} )
+  \frac{4 \pi }{R_1 (t)} \delta (\vec R_2).
\label{Efielddiv09}
\enq
The first term in the right hand side is a divergence. Thus using Gauss theorem its volume integral will become a surface integral, the second term is a delta function. This means that there is no contribution to the volume integral from that terms unless $\vec x = \vec x_2$. We may now write \ern{Efielddiv06} as follows:
\beq
E'^{[0]}_{field~12} =   I_1 (t)  I_2    \oint \oint (d \tilde{\vec{l_1}} \cdot d \tilde{\vec{l_2}}) [(\int d a \hat n
\cdot \frac{1}{R_1 (t)}   \vec \nabla \frac{1}{R_2 (t)})+\frac{4 \pi }{R_{12} (t)} ].
\label{Efielddiv10}
\enq
Let us look at the surface integral and assume that the system is contained inside an infinite sphere. On the surface of such a sphere:
\beq
d a \hat n  = r^2 d \Omega \hat r, \qquad \hat r =\frac{\vec x}{r}
\label{spherele}
\enq
$d \Omega$ is an infinitesimal solid angle. Now since $r=|\vec x| \rightarrow \infty$ on the surface and:
\beq
\vec R = \vec x - \vec x' = r (\hat r  - \frac{\vec x'}{r})
\label{asym1}
\enq
it follows that:
\beq
R = |\vec x - \vec x'| = r |\hat r  - \frac{\vec x'}{r}| \simeq r (1 - \hat r \cdot \frac{\vec x'}{r}) = r - \hat r \cdot \vec x'
\label{asym2}
\enq
up to second order in the infinitesimal quantity $\frac{\vec x'}{r}$. And for the same reason:
\beq
\frac{1}{R}  \simeq  \frac{1}{r} (1 + \hat r \cdot \frac{\vec x'}{r})
\label{asym3}
\enq
to the same order. And also:
\beq
R^3 \simeq r^3 (1 - 3 \hat r  \cdot \frac{\vec x'}{r}), \qquad \frac{1}{R^3} \simeq \frac{1}{r^3} (1 + 3 \hat r  \cdot \frac{\vec x'}{r})
\label{asym4}
\enq
\beq
 \vec \nabla \frac{1}{R}   = - \frac{\vec R }{R^3} \simeq - \frac{1}{r^2} [(1 + 3 \hat r  \cdot \frac{\vec x'}{r})\hat r -\frac{\vec x'}{r})]
 \label{asym5}
\enq
Using the above results we conclude that:
\beq
\lim_{r \rightarrow \infty}  \int d a \hat n \cdot \frac{1}{R_1 (t)} \vec \nabla \frac{1}{R_2 (t)} =
 -\lim_{r \rightarrow \infty}   \int d \Omega  \frac{1}{r } = -\lim_{r \rightarrow \infty} \frac{4 \pi}{r } =0
  \label{asym6}
\enq
Hence:
\beq
E'^{[0]}_{field~12} =  I_1 (t)  I_2  \oint \oint (d \tilde{\vec{l_1}} \cdot d \tilde{\vec{l_2}}) [\frac{4 \pi }{R_{12} (t)}].
\label{Efielddiv11}
\enq
And:
\beq
E_{field~12}^{[0]} =  \frac{\mu_0}{(4 \pi)^2} E'^{[0]}_{field~12}   = I_1 (t)  I_2  M^{[0]}_{12}.
\label{Efielddiv12}
\enq
where we took advantage of the mutual inductance term defined in \ern{mutind}.

\subsubsection{Poynting vector}

Finally we shall study the Poynting vector:
\beq
\vec S'^{[0]}_{p~12} = \vec E'^{[0]}_1 \times \vec B'^{[0]}_2 +\vec E'^{[0]}_2 \times \vec B'^{[0]}_1
\label{Poyntingv0}
\enq
Taking into account that $\vec E_2^{[0]}$ is null according to \ern{E02} this simplifies to:
\beq
\vec S'^{[0]}_{p~12} = \vec E'^{[0]}_1 \times \vec B'^{[0]}_2.
\label{Poyntingv0b}
\enq
Now the Poynting vector terms which is usually associated with radiation only contributed to the energy balance on a surface
encapsulating the system under consideration (see \ern{Econt0pn}). Taking this surface to be spherical and at infinity we deduce
that only the asymptotic forms of $\vec E'^{[0]}_1$ and $\vec B'^{[0]}_2$ are of interest for the purpose of evaluating the Poynting vector
contribution.  According to \ern{E0}:
\beq
\vec E'^{[0]} = -  \partial_t I (t)  \oint d \tilde{\vec{l}} \frac{1}{R(t)} \simeq
-  \partial_t I (t) \frac{1}{r^2} \oint d \vec{l} \  \hat r \cdot \vec x'
\label{E0b2}
\enq
In which we use the approximation given in \ern{asym3} and take into account that $\oint d \tilde{\vec{l}} =0$. The magnetic field is given
in \ern{B0}:
\beq
\vec B'^{[0]} (\vec x,t) = I (t)  \oint d \tilde{\vec{l}} \times  \frac{\vec R(t) }{R^3 (t)} \simeq
 I (t) \frac{1}{r^3} \oint d \vec{l} \times [ 3 (\hat r  \cdot \vec x') \hat r -\vec x'].
\label{B0b}
\enq
in the above we used again the identity $\oint d \vec l =0$. We shall define the quantity:
\beq
\vec \Lambda = \oint d \vec{l} \times [ 3 (\hat r  \cdot \vec x') \hat r -\vec x'].
\label{Lam}
\enq
for future use.  From the above it easy to see that:
\beq
\lim_{r \rightarrow \infty} \vec S'^{[0]}_{p~12} \propto \frac{1}{r^5}
\enq
and thus it easy to see that:
\beq
\oint_S \vec S'^{[0]}_{p~12} \cdot \hat n da = \lim_{r \rightarrow \infty} \oint_S (\vec S'^{[0]}_{p~12} \cdot \hat r) r^2 d \Omega  = 0
\enq
hence there is no Poynting vector contribution to the energy balance. Thus is the quasi static approximation there is no radiation losses as expected.

\subsubsection{Intermediate account}

We conclude that the energy equation of order zero is indeed balanced. Mechanical work invested or extracted in the system results in increase or decrease in the field energy accordingly. To be more specific the magnetic field energy is affected by the mechanical work.
The power related to the mechanical work is:
\beq
Power_{12}^{[0]} = -  \partial_t I_1 (t) I_2  M^{[0]}_{12}
\label{pow05b}
\enq
and this is equal to minus the derivative of the field energy:
\beq
E_{field~12}^{[0]}  = I_1 (t)  I_2  M^{[0]}_{12}.
\label{Efielddiv12b}
\enq
which is what is expected from \ern{Econt0pn} for the case that there is no Poynting contribution. We underline that those contributions are not related the relativistic engine effect as non of the terms depends on the engine velocity $\vec v$, and thus the above expression will be valid even if the egnine is infinitely massive and no motion occurs.

\subsection{$n=1$}

Let us look at \ern{Econt0pn} and study it for the first order in $\frac{1}{c}$:
\beq
Power'^{[1]}_{12} = -\frac{d  E'^{[1]}_{field~12}}{dt} -\oint_S \vec S'^{[1]}_{p~12} \cdot \hat n da.
\label{Econt0ord1}
\enq

\subsubsection{Power}

We shall start by calculating $Power_{12}^{[1]}$, according to \ern{Powerpartp2lo}:
\beq
Power'^{[1]}_{12} = 4 \pi   \left( I_1 (t) \oint d \tilde{\vec{l_1}} \cdot \vec E'^{[1]}_2 (\vec x_1) + I_2 (t) \oint d \tilde{\vec{l_2}} \cdot \vec E'^{[1]}_1 (\vec x_2) \right).
\label{pow1}
\enq
The field $\vec E'^{[1]}$ can only come from the vector potential $\vec A'^{[1]}$. As $\vec A'^{(1)}=0$  (see \ern{A1null})
it follows that $\vec A'^{[1]} = 0$ and hence by \ern{E1} :
\beq
\vec E'^{[1]} = 0.
\label{Eo1}
\enq
I then follows that:
\beq
Power'^{[1]}_{12} = 0.
\label{pow12}
\enq
Hence no mechanical work is invested nor extracted for a relativistic engine in the first order of $\frac{1}{c}$.

\subsubsection{Field Energy}

Turning our attention next to field energy defined in \ern{Efielddivp2}:
\beq
E'^{[1]}_{field~12} = \int [\vec B'^{[0]}_1 \cdot \vec B'^{[1]}_2  + \vec B'^{[1]}_1 \cdot \vec B'^{[0]}_2 ] d^3 x.
\label{Efielddiv1}
\enq
$\vec B'^{[1]}$ is calculated according to \ern{Bgeneraln} as:
\beq
\vec B'^{[1]} = \vec \nabla \times \vec A'^{[1]}.
\label{Bgeneraln1}
\enq
Now since $ \vec A'^{[1]}= 0 $ we may use \ern{Bgeneraln1} to obtain:
\beq
\vec B'^{[1]} (\vec x,t) = 0.
\label{B1}
\enq
for the magnetic field generated from both coils. Inserting \ern{B1} into \ern{Efielddiv1} will yield:
\beq
E'^{[1]}_{field~12} =  0.
\label{Efielddiv13}
\enq
Hence there is no contribution from first order terms in $\frac{1}{c}$ to the field energy of a relativistic engine neither.

\subsubsection{Poynting vector}

Finally we shall study the Poynting vector:
\beq
\vec S'^{[1]}_{p~12} = \vec E'^{[0]}_1 \times \vec B'^{[1]}_2 +\vec E'^{[1]}_1 \times \vec B'^{[0]}_2 +\vec E'^{[0]}_2 \times \vec B'^{[1]}_1
+\vec E'^{[1]}_2 \times \vec B'^{[0]}_1
\label{Poyntingv1}
\enq
According to \ern{Eo1} and \ern{B1} all the above terms vanish and we have:
\beq
\vec S'^{[1]}_{p~12} = 0 \Rightarrow \oint_S \vec S'^{[1]}_{p~12} \cdot \hat n da =0.
\label{Poyntingv1b}
\enq

\subsubsection{Intermediate account}

We conclude that the energy equation of order one is indeed balanced in a trivial way.
\Ern{Econt0pn} is satisfied in the sense that $0=0$. From now on we will disregard in our calculations any field term of order one, as their contribution to any expression is obviously null.

\subsection{$n=2$}

Let us look at \ern{Econt0pn} and study it for the second order in $\frac{1}{c}$:
\beq
Power'^{[2]}_{12} = -\frac{d  E'^{[2]}_{field~12}}{dt} -\oint_S \vec S'^{[2]}_{p~12} \cdot \hat n da.
\label{Econt0ord2}
\enq

\subsubsection{Power}

We shall start by calculating $Power_{12}^{[2]}$, according to \ern{Powerpartp2lo}:
\beq
Power'^{[2]}_{12} = 4 \pi   \left( I_1 (t) \oint d \tilde{\vec{l_1}} \cdot \vec E'^{[2]}_2 (\vec x_1) + I_2 (t) \oint d \tilde{\vec{l_2}} \cdot \vec E'^{[2]}_1 (\vec x_2) \right).
\label{pow2}
\enq
The field $\vec E'^{[2]}$ can only come from the vector potential $\vec A'^{[2]}$ and $\vec A'^{[0]}$. $\vec A'^{[0]}$ is given in
\ern{Ab0} while $\vec A'^{[2]}$ can be deduce from  $\vec A'^{(2)}$ which can be calculated using \ern{A4b} as:
\beq
\vec A'^{(2)} (\vec x,t) =  \frac{1}{2 c^2} \frac{d^2}{dt^2} \left[ I (t) \oint d \tilde{\vec{l}} R(t) \right] ,
\label{A4b2}
\enq
Or explicitly as:
\ber
\vec A'^{(2)} (\vec x,t) &=&  \frac{1}{2 c^2} \left\{ \frac{d^2 I (t)}{dt^2} \oint d \tilde{\vec{l}}R(t)
-2 \frac{d I(t)}{dt} \oint d \tilde{\vec{l}} \hat R \cdot \vec v \right.
\nonumber \\
&+&\left. I(t) \oint d \tilde{\vec{l}} \left[ \frac{1}{R}(\hat R \times \vec v)^2 - \hat R \cdot \frac{d \vec v}{dt} \right] \right\},
\label{A4b2b}
\enr
in which we have used \ern{dR0}. It is clear that $\vec A'^{(2)}$ contain contributions to $\vec A'^{[2]}$ but also to $\vec A'^{[4]}$
and $\vec A'^{[6]}$ (but not to odd $\vec A'^{[n]}$'s). For now we will only be interested in $\vec A'^{[2]}$ which is:
\beq
\vec A'^{[2]} (\vec x,t) \equiv \frac{1}{2 c^2}  \frac{d^2 I (t)}{dt^2} \oint d \tilde{\vec{l}}R(t).
\label{Asec2}
\enq
Combining the above result with \ern{E1b} and \ern{E0b} yields:
\beq
\vec E'^{[2]} =  - \frac{1}{2 c^2}  \frac{d^3 I (t)}{dt^3} \oint d \tilde{\vec{l}}R(t) -
I (t) \oint d \tilde{\vec{l}}\quad \frac{\vec v \cdot \vec R(t)}{R^3(t)}.
\label{E2b}
\enq
Now since the second coil has a constant current this leads to the result:
\beq
\vec E'^{[2]}_2 = - I_2 \oint d \tilde{\vec{l_2}}\quad \frac{\vec v \cdot \vec R_2 (t)}{R_2^3(t)}.
\label{E22}
\enq
Inserting the above results into \ern{pow2} yields:
\ber
Power'^{[2]}_{12} = &-& 4 \pi I_1(t) I_2 \oint \oint d \tilde{\vec{l_1}} \cdot d \tilde{\vec{l_2}} \frac{\vec v \cdot \vec R_{12}} {R_{12}^3}
\nonumber \\
&-& 4 \pi I_1(t) I_2 \oint \oint d \tilde{\vec{l_2}} \cdot d \tilde{\vec{l_1}} \frac{\vec v \cdot \vec R_{21}}{R_{21}^3}
\nonumber \\
&-& \frac{2 \pi}{c^2} \frac{d^3 I_1 (t)}{dt^3}  I_2 \oint \oint d \tilde{\vec{l_1}} \cdot d \tilde{\vec{l_2}} R_{12}
\label{pow22}
\enr
In the above:
\beq
\vec R_{21} = \vec x_2 - \vec x_1 = - \vec R_{12} \Rightarrow R_{21} = |\vec R_{21}| = |\vec R_{12}| = R_{12}
\label{pow23}
\enq
Inserting \ern{pow23} into \ern{pow22} will result in the cancellation of the first two terms:
\ber
Power'^{[2]}_{12} = - \frac{2 \pi}{c^2} \frac{d^3 I_1 (t)}{dt^3}  I_2 \oint \oint d \vec{l_2} \cdot d \vec{l_2} R_{12}
\label{pow24}
\enr
Hence:
\beq
Power_{12}^{[2]} = \frac{\mu_0}{16 \pi^2} Power'^{[2]}_{12} = - \frac{\mu_0}{8 \pi c^2} \frac{d^3 I_1 (t)}{dt^3}  I_2
\oint \oint d \vec{l_2} \cdot d \vec{l_2} R_{12}
\label{pow25}
\enq
We notice that $Power_{12}^{[2]}$ may be positive or negative according to the relative position of the current loops and current directions
and third derivative, hence, energy can be invested or extracted from the combined system according to the system configuration. We also
notice that this term has nothing to do with the relativistic engine effect as it is completely independent of the mass of the engine and will exist also for an infinitely heavy motor. This is to be expected as the relativistic engine effect is fourth order in $\frac{1}{c}$ and not second order. We will interpret the double integral $\oint \oint d \vec{l_2} \cdot d \vec{l_2} R_{12}$ in the next subsection dealing with the field energy.

\subsubsection{Field Energy}

Turning our attention next to field energy defined in \ern{Efielddivp2} we obtain the following expression for second order terms in $\frac{1}{c}$:
\beq
E'^{[2]}_{field~12} = \int \left( \frac{1}{c^2} \vec E'^{[0]}_1 \cdot \vec E'^{[0]}_2 +  \vec B'^{[0]}_1 \cdot \vec B'^{[2]}_2
+  \vec B'^{[2]}_1 \cdot \vec B'^{[0]}_2\right)   d^3 x,
\label{Efielddiv2}
\enq
in which we are reminded that there are no field contributions which are first order in $\frac{1}{c}$.
According to \ern{E02} the zeroth order electric field for the static current second coil is null, hence:
\beq
E'^{[2]}_{field~12} = \int \left( \vec B'^{[0]}_1 \cdot \vec B'^{[2]}_2 +  \vec B'^{[2]}_1 \cdot \vec B'^{[0]}_2\right)   d^3 x,
\label{Efielddiv2b}
\enq
$\vec B'^{[2]}$ is calculated according to \ern{Bgeneraln} as:
\beq
\vec B'^{[2]} = \vec \nabla \times \vec A'^{[2]}.
\label{Bgeneraln2}
\enq
Taking into account \ern{Asec2} the second order correction to the magnetic field is thus:
\beq
\vec B'^{[2]} (\vec x,t) = - \frac{1}{2 c^2}  \frac{d^2 I (t)}{dt^2} \oint d \tilde{\vec{l}} \times \vec \nabla R (t)
= - \frac{1}{2 c^2}  \frac{d^2 I (t)}{dt^2} \oint d \tilde{\vec{l}} \times \frac{\vec R(t)}{R(t)}.
\label{Bsec2}
\enq
Hence for a static coil:
\beq
\vec B'^{[2]}_2 (\vec x,t) = 0.
\label{Bsec2b}
\enq
And thus:
\beq
E'^{[2]}_{field~12} = \int  \vec B'^{[2]}_1 \cdot \vec B'^{[0]}_2 d^3 x.
\label{Efielddiv2c}
\enq
Inserting \ern{B0} into \ern{Efielddiv0} will yield:
\beq
E'^{[2]}_{field~12} =   \frac{1}{2 c^2}  \frac{d^2 I_1 (t)}{dt^2}  I_2   \int d^3 x  \oint d \tilde{\vec{l_1}} \times  \vec \nabla R_1 (t) \cdot \oint d \tilde{\vec{l_2}} \times  \vec \nabla \frac{1}{R_2 (t)}
\label{Efielddiv23}
\enq
in which we recall that $I_2$ is time independent. Using a well known identity from vector analysis we may write:
\ber
E'^{[2]}_{field~12} &=&   \frac{1}{2 c^2}  \frac{d^2 I_1 (t)}{dt^2}   I_2   \int d^3 x  \oint \oint [
(d \tilde{\vec{l_1}} \cdot d \tilde{\vec{l_2}}) (\vec \nabla R_1 (t) \cdot  \vec \nabla \frac{1}{R_2 (t)})
\nonumber \\
 & - & (d \tilde{\vec{l_1}} \cdot \vec \nabla \frac{1}{R_2 (t)}) (d \tilde{\vec{l_2}} \cdot \vec \nabla R_1 (t)) ].
\label{Efielddiv24}
\enr
Let us look at the integral expression
\beq
int_2=\int d^3 x  \oint \oint [ (d \tilde{\vec{l_1}} \cdot \vec \nabla \frac{1}{R_2 (t)}) (d \tilde{\vec{l_2}} \cdot \vec \nabla R_1 (t)) ]
\label{Efielddiv25}
\enq
This is an expression of the type described in \ern{Efielddiv05d} of appendix A with $h =\frac{1}{R_2}$  and  $g = R_1$.
According to appendix A the expression in \ern{Efielddiv25} can be expressed as a surface integral. Assuming that
our system is contained in an infinite sphere we have according to \ern{Efielddiv05gd} and \ern{spherele}:
\beq
int_2 =\oint \oint d  l_{1n} d  l_{2m}  \lim_{r \rightarrow \infty} \oint d \Omega \ r^2 \hat r_n  \partial_m R_1 \frac{1}{R_2}
\label{Efielddiv05gdor2}
\enq
Now:
\beq
\partial_m R_1 = \frac{\vec R_{1m}}{R_1} = \hat R_{1m}
\label{parmR}
\enq
This can be calculated up to second order in $\frac{x_1}{r}$ using \ern{asym1} and \ern{asym2} as:
\beq
\hat R_{1m} = \hat r_m + \frac{1}{r}(\hat r_m(\hat r \cdot \vec x_1) - \vec x_{1m}) + O \left((\frac{x_1}{r})^2\right).
\label{hatRa}
\enq
Similarly according to \ern{asym3}:
\beq
\frac{1}{R}  \simeq  \frac{1}{r} \left[1 +  \frac{\hat r \cdot \vec x_2}{r}+ O \left((\frac{x_2}{r})^2\right)\right].
\label{asym3b}
\enq
Plugging \ern{hatRa} and \ern{asym3b} into \ern{Efielddiv05gdor2} we obtain:
\ber
& & \hspace{-1cm} int_2 = \oint \oint d  l_{1n} d  l_{2m}  \lim_{r \rightarrow \infty} \oint d \Omega \ r \hat r_n
\nonumber \\
& & \hspace{-1cm}
\left[ \hat r_m + \frac{1}{r}(\hat r_m(\hat r \cdot \vec x_1) - \vec x_{1m}) + O \left((\frac{x_1}{r})^2\right) \right]
 \left[1 +  \frac{\hat r \cdot \vec x_2}{r}+ O \left((\frac{x_2}{r})^2\right)\right]
\label{Efielddiv05gdor333}
\enr
The $O \left((\frac{x'}{r})^2\right)$ terms will not contribute in the limit of infinite radius and thus we may write:
\beq
  int_2 = \oint \oint d  l_{1n} d  l_{2m}  \lim_{r \rightarrow \infty} \oint d \Omega \ r \hat r_n
\left[ \hat r_m + \frac{1}{r}(\hat r_m(\hat r \cdot \vec x_1) - \vec x_{1m}) \right]
 \left[1 +  \frac{\hat r \cdot \vec x_2}{r}\right]
\label{Efielddiv05gdor4}
\enq
multiplying the square brackets explicitly we obtain:
\ber
int_2 &=& \oint \oint d  l_{1n} d  l_{2m}  \lim_{r \rightarrow \infty} \oint d \Omega \ r \hat r_n
\left[ \hat r_m + \frac{1}{r}(\hat r_m(\hat r \cdot (\vec x_1 +\vec x_2) ) - \vec x_{1m}) \right.
\nonumber \\
&+& \left. O \left(\frac{ x_1 x_2}{r^2}\right) \right]
 \label{Efielddiv05gdor5}
\enr
in the limit of large $r$:
\beq
int_2 = \oint \oint d  l_{1n} d  l_{2m}  \lim_{r \rightarrow \infty} \oint d \Omega \ r \hat r_n
\left[ \hat r_m + \frac{1}{r}(\hat r_m(\hat r \cdot (\vec x_1 +\vec x_2) ) - \vec x_{1m}) \right]
 \label{Efielddiv05gdor6}
\enq
Now we notice that performing a loop integral over a constant $\vec C$ we obtain a null result:
\beq
\oint d  \vec l \cdot \vec C = 0
 \label{lopC}
\enq
Since neither of the terms in the square bracket depends on both $\vec x_1$ and  $\vec x_2$ (they depend on either $\vec x_1$ or $\vec x_2$ or non of them) certainly one of the loop integrals of \ern{Efielddiv05gdor6} will vanish (or both), hence $int_2=0$ and thus
\ern{Efielddiv24} takes the form:
\beq
E'^{[2]}_{field~12} =   \frac{1}{2 c^2}  \frac{d^2 I_1 (t)}{dt^2}  I_2   \int d^3 x  \oint \oint [
(d \tilde{\vec{l_1}} \cdot d \tilde{\vec{l_2}}) (\vec \nabla R_1 (t) \cdot  \vec \nabla \frac{1}{R_2 (t)})].
\label{Efielddiv26}
\enq
Now:
\beq
\vec \nabla R_1 (t)  \cdot  \vec \nabla \frac{1}{R_2 (t)} = \vec \nabla \cdot (R_1 (t)  \vec \nabla \frac{1}{R_2 (t)} )
- R_1 (t) \vec \nabla^2 \frac{1}{R_2 (t)}
\label{Efielddiv27}
\enq
Taking into account \ern{Efielddiv08}, we have:
\beq
\vec \nabla R_1 (t) \cdot  \vec \nabla \frac{1}{R_2 (t)} = \vec \nabla \cdot (R_1 (t) \vec \nabla \frac{1}{R_2 (t)} )
+  4 \pi R_1 (t) \delta (\vec R_2).
\label{Efielddiv29}
\enq
The first term in the right hand side is a divergence. Thus using Gauss theorem its volume integral will become a surface integral, the second term is a delta function. This means that there is no contribution to the volume integral from that terms unless $\vec x = \vec x_2$. We may now write \ern{Efielddiv26} as follows:
\beq
E'^{[2]}_{field~12} =   \frac{1}{2 c^2}  \frac{d^2 I_1 (t)}{dt^2}   I_2    \oint \oint (d \tilde{\vec{l_1}} \cdot d \tilde{\vec{l_2}}) [\int d a \hat n \cdot R_1 (t) \vec \nabla \frac{1}{R_2 (t)} +  4 \pi R_{12} (t) ].
\label{Efielddiv210}
\enq
Let us look at the surface integral and assume as usual that the system is contained inside an infinite sphere.
\ber
int_3 &=& \oint \oint (d \tilde{\vec{l_1}} \cdot d \tilde{\vec{l_2}}) \int d a \hat n \cdot R_1 (t) \vec \nabla \frac{1}{R_2 (t)}
\nonumber \\
& = &
\oint \oint (d \tilde{\vec{l_1}} \cdot d \tilde{\vec{l_2}}) \lim_{r \rightarrow \infty} \int d \Omega \ r^2 \hat r \cdot R_1 (t) \vec \nabla \frac{1}{R_2 (t)}
\label{Efielddiv211}
\enr
Using \ern{asym2} and \ern{asym5} this can be written as:
\ber
int_3 & = &
-\oint \oint (d \tilde{\vec{l_1}} \cdot d \tilde{\vec{l_2}}) \lim_{r \rightarrow \infty} \int d \Omega \ r^2 \hat r \cdot r \left[1 - \hat r \cdot \frac{\vec x_1}{r}+O \left((\frac{x_1}{r})^2\right)\right]
\nonumber \\
& &
\frac{1}{r^2} \left[\hat r + \frac{1}{r}(3 (\hat r  \cdot \vec x_2) \hat r - \vec x_2)+ O \left((\frac{x_2}{r})^2\right) \right]
\label{Efielddiv212}
\enr
Or also as:
\ber
int_3 & = &
-\oint \oint (d \tilde{\vec{l_1}} \cdot d \tilde{\vec{l_2}}) \lim_{r \rightarrow \infty} \int d \Omega \  r \left[1 - \hat r \cdot \frac{\vec x_1}{r}+O \left((\frac{x_1}{r})^2\right)\right]
\nonumber \\
& &
 \left[1 + \frac{2}{r} (\hat r  \cdot \vec x_2) + O \left((\frac{x_2}{r})^2\right) \right]
\label{Efielddiv213}
\enr
In the limit of large $r$ the terms $O \left((\frac{x}{r})^2\right)$ will not contribute, hence:
\beq
int_3 =
-\oint \oint (d \tilde{\vec{l_1}} \cdot d \tilde{\vec{l_2}}) \lim_{r \rightarrow \infty} \int d \Omega \  r \left[1 - \hat r \cdot \frac{\vec x_1}{r}\right]  \left[1 + \frac{2}{r} (\hat r  \cdot \vec x_2) \right]
\label{Efielddiv214}
\enq
multiplying the square brackets explicitly we obtain:
\beq
int_3 =
-\oint \oint (d \tilde{\vec{l_1}} \cdot d \tilde{\vec{l_2}}) \lim_{r \rightarrow \infty} \int d \Omega \  r \left[1  + \frac{1}{r} (2 \hat r  \cdot \vec x_2 - \hat r  \cdot \vec x_1 ) + O \left((\frac{ x_1 x_2}{r^2})\right) \right]
\label{Efielddiv215}
\enq
the term $O \left((\frac{ x_1 x_2}{r^2})\right)$ will not contribute, hence:
\beq
int_3 =
-\oint \oint (d \tilde{\vec{l_1}} \cdot d \tilde{\vec{l_2}}) \lim_{r \rightarrow \infty} \int d \Omega \  r \left[1  + \frac{1}{r} (2 \hat r  \cdot \vec x_2 - \hat r  \cdot \vec x_1 ) \right] = 0
\label{Efielddiv216}
\enq
Since neither of the terms in the square bracket depends on both $\vec x_1$ and  $\vec x_2$ (they depend on either $\vec x_1$ or $\vec x_2$ or non of them) certainly one of the loop integrals of \ern{Efielddiv05gdor6} will vanish (or both) being a loop integral
over a constant vector (see \ern{lopC}), hence $int_3=0$ and thus \ern{Efielddiv210} takes the simple form:
\beq
E'^{[2]}_{field~12} =   \frac{2 \pi}{c^2}  \frac{d^2 I_1 (t)}{dt^2}   I_2    \oint \oint (d \vec{l_1} \cdot d \vec{l_2}) R_{12}.
\label{Efielddiv217}
\enq
And:
\beq
E_{field~12}^{[2]} =  \frac{\mu_0}{(4 \pi)^2} E'^{[2]}_{field~12}   =  \frac{\mu_0}{8 \pi c^2}  \frac{d^2 I_1 (t)}{dt^2}   I_2    \oint \oint (d \vec{l_1} \cdot d \vec{l_2}) R_{12}.
\label{Efielddiv218}
\enq
For a phasor current with frequency $\omega$:
\beq
I_1 (t) = I_{1 0} e^{j \omega t}, \qquad j \equiv \sqrt{-1}
\label{phas}
\enq
we obtain a second order correction to the mutual inductance of the form \cite{{RY}}:
\beq
M^{[2]}_{12} \equiv -\frac{\mu_0 \omega^2}{8 \pi c^2} \oint d \vec l_1 \cdot \oint d \vec l_2 R_{12}
\label{CI12}
\enq
Such that:
\beq
E_{field~12}^{[2]} =  I_1 (t)  I_2  M^{[2]}_{12}.
\label{Efielddiv219}
\enq
Obviously the larger the system and the higher the frequency the more important this correction is. We stress that
this term is not related to the relativistic engine effect and will exist even for an engine of "infinite" mass.

\subsubsection{Poynting vector}

Finally we shall study the Poynting vector:
\beq
\vec S'^{[2]}_{p~12} = \vec E'^{[0]}_1 \times \vec B'^{[2]}_2 + \vec E'^{[2]}_1 \times \vec B'^{[0]}_2 +
\vec E'^{[0]}_2 \times \vec B'^{[2]}_1 + \vec E'^{[2]}_2 \times \vec B'^{[0]}_1
\label{Poyntingv2}
\enq
Taking into account that $\vec E_2^{[0]}$ is null according to \ern{E02} and $ \vec B'^{[2]}_2 $  is null according to \ern{Bsec2b}
this simplifies to:
\beq
\vec S'^{[2]}_{p~12} = \vec E'^{[2]}_1 \times \vec B'^{[0]}_2 +  \vec E'^{[2]}_2 \times \vec B'^{[0]}_1
\label{Poyntingv2b}
\enq
Now the Poynting vector terms which is usually associated with radiation only contributes to the energy balance on a surface
encapsulating the system under consideration (see \ern{Econt0pn}). Taking this surface to be spherical and at infinity we deduce
that only the asymptotic forms of $\vec E'^{[2]}$ and $\vec B'^{[0]}$ are of interest for the purpose of evaluating the Poynting vector
contribution.  According to \ern{E2b}:
\beq
\vec E'^{[2]} =  - \frac{1}{2 c^2}  \frac{d^3 I (t)}{dt^3} \oint d \tilde{\vec{l}}R(t) -
I (t) \oint d \tilde{\vec{l}}\quad \frac{\vec v \cdot \vec R(t)}{R^3(t)}.
\label{E2bc}
\enq
Taking into account the asymptotic forms given in \ern{asym2} and \ern{asym5}
and taking into account that $\oint d \tilde{\vec{l}} =0$., we arrive at the asymptotic form:
\beq
\vec E'^{[2]} \simeq   \frac{1}{2 c^2}  \frac{d^3 I (t)}{dt^3} \oint d \vec{l} \ \hat n \cdot \vec x' +
 \frac{I (t)}{r^3}\oint d \vec{l} [\vec v \cdot \vec x' - 3(\vec v \cdot \hat n) (\hat n \cdot \vec x')] .
\label{E2bas}
\enq
Hence for an asymptotic field created by a time dependent current we have the form :
\beq
\vec E'^{[2]}_1 \simeq   \frac{1}{2 c^2}  \frac{d^3 I_1 (t)}{dt^3} \oint d \vec{l_1} \ \hat n \cdot \vec x_1 .
\label{E2bas1}
\enq
However, for an asymptotic field created by a time independent current we have the form :
\beq
\vec E'^{[2]}_2 \simeq   \frac{I_2}{r^3}\oint d \vec{l_2} [\vec v \cdot \vec x_2 - 3(\vec v \cdot \hat n) (\hat n \cdot \vec x_2)]
\label{E2bas2}
\enq
The asymptotic form of the magnetic field $\vec B'^{[0]}$ is given in \ern{B0b} and \ern{Lam}:
\beq
\vec B'^{[0]} (\vec x,t)  \simeq   \frac{I (t) \vec \Lambda}{r^3}.
\label{B0bc}
\enq
From the above it easy to see that:
\beq
\lim_{r \rightarrow \infty} \vec S'^{[2]}_{p~12} \propto \frac{1}{r^3}
\enq
of course the contribution from \ern{E2bas2} will go to zero much faster as: $\frac{1}{r^6}$.
It easy to see that:
\beq
\oint_S \vec S'^{[2]}_{p~12} \cdot \hat n da = \lim_{r \rightarrow \infty} \oint_S (\vec S'^{[2]}_{p~12} \cdot \hat r) r^2 d \Omega  = 0
\enq
hence there is no Poynting vector contribution to the energy balance. This is expected as the mechanical work is balanced by the field energy loss exactly in second order terms of $\frac{1}{c}$ .

\subsubsection{Intermediate account}

We conclude that the energy equation of the second order is indeed balanced. Mechanical work invested or extracted in the system results in increase or decrease in the field energy accordingly. To be more specific the magnetic field energy is affected by the mechanical work.
The power related to the mechanical work is according to \ern{pow25}:
\beq
Power_{12}^{[2]} =  - \frac{\mu_0}{8 \pi c^2} \frac{d^3 I_1 (t)}{dt^3}  I_2 \oint \oint d \vec{l_2} \cdot d \vec{l_2} R_{12}
\label{pow25b}
\enq
and this is equal to minus the derivative of the field energy \ern{Efielddiv218}:
\beq
E_{field~12}^{[2]} =  \frac{\mu_0}{8 \pi c^2}  \frac{d^2 I_1 (t)}{dt^2}   I_2    \oint \oint (d \vec{l_1} \cdot d \vec{l_2}) R_{12}.
\label{Efielddiv218b}
\enq
which is what is expected from \ern{Econt0ord2} for the case that there is no Poynting contribution. We underline that those contributions are not related the relativistic engine effect as non of the terms depends on the engine velocity $\vec v$, and thus the above expression will be valid even if the engine is infinitely massive and no motion occurs. We do not expect any relativistic engine contributions for orders smaller than $\frac{1}{c^4}$.
For a phasor current of frequency $\omega$ \ern{CI12} indicates a relativistic correction to the classical mutual inductance which is important for large systems with high frequency.

\subsection{$n=3$}

Let us look at \ern{Econt0pn} and study it for the third order in $\frac{1}{c}$:
\beq
Power'^{[3]}_{12} = -\frac{d  E'^{[3]}_{field~12}}{dt} -\oint_S \vec S'^{[3]}_{p~12} \cdot \hat n da.
\label{Econt0ord3}
\enq

\subsubsection{Power}

We shall start by calculating $Power_{12}^{[3]}$, according to \ern{Powerpartp2lo}:
\beq
Power'^{[3]}_{12} = 4 \pi   \left( I_1 (t) \oint d \tilde{\vec{l_1}} \cdot \vec E'^{[3]}_2 (\vec x_1) + I_2 (t) \oint d \tilde{\vec{l_2}} \cdot \vec E'^{[3]}_1 (\vec x_2) \right).
\label{pow3}
\enq
The field $\vec E'^{[3]}$ can only come from the vector potential $\vec A'^{[3]}$ and $\vec A'^{[1]}$. $\vec A'^{[1]}$ is null according to
\ern{A1null} while $\vec A'^{[3]}$ can be deduce from  $\vec A'^{(3)}$ which can be calculated using \ern{A4b} as:
s expected the higher the frequency the more pronounced the relativistic effect is. A third order correction to the magnetic
field in a relativistic motor can be derived from the third term $\vec A'^{(3)}$  in the sum given in \ern{A4}:
\beq
\vec A'^{(3)} (\vec x,t) = - \frac{1}{6  c^3}  \frac{d^3}{dt^3} \left[ I (t) \oint d \tilde{\vec{l}}R^2(t) \right].
\label{Athird}
\enq
Now it is clear that this expression contains terms of the order of $\frac{1}{c^3}$ and higher (including terms of the order $\frac{1}{c^5}$
and $\frac{1}{c^7}$). However, it is easy to see that there are no third order corrections to the vector potential except:
\beq
\vec A'^{[3]} (\vec x,t) = - \frac{1}{6  c^3}   \frac{d^3 I (t)}{dt^3} \oint d \tilde{\vec{l}}R^2(t).
\label{Athir2}
\enq
Now using  \ern{E1} we arrive at the following equation for $\vec E^{[3]}$:
\beq
\vec E'^{[3]} = \frac{1}{6 c^3} \frac{d^4 I (t)}{d t^4} \oint d \vec l R^2
 \label{Eb3}
\enq
in which we maintained only terms of the order of $\frac{1}{c^3}$ but not higher.
From which it is clear that:
\beq
E'^{[3]}_2=0
\label{Eb3b}
\enq
for a constant current loop. Hence:
\beq
Power'^{[3]}_{12} = 4 \pi I_2 \oint d \tilde{\vec{l_2}} \cdot \vec E'^{[3]}_1 (\vec x_2) .
\label{pow3b}
\enq
\beq
Power'^{[3]}_{12} = \frac{2 \pi}{3 c^3} \frac{d^4 I_1 (t)}{dt^4}  I_2 \oint \oint d \tilde{\vec{l_1}} \cdot d \tilde{\vec{l_2}} R^2_{12}
\label{pow32}
\enq
Hence:
\beq
Power_{12}^{[3]} = \frac{\mu_0}{16 \pi^2} Power'^{[3]}_{12} =  \frac{\mu_0}{24 \pi c^3} \frac{d^4 I_1 (t)}{dt^4}  I_2
\oint \oint d \vec{l_1} \cdot d \vec{l_2} R^2_{12}
\label{pow35}
\enq
We notice that $Power_{12}^{[3]}$ may be positive or negative according to the relative position of the current loops and current directions
and fourth derivative, hence, energy can be invested or extracted from the combined system according to the system configuration. We also
notice that this term has nothing to do with the relativistic engine effect as it is completely independent of the mass of the engine and will exist also for an infinitely heavy motor. This is to be expected as the relativistic engine effect is fourth order in $\frac{1}{c}$ and not third order. We will interpret the double integral $\oint \oint d \vec{l_1} \cdot d \vec{l_2} R^2_{12}$ in the next subsection dealing with the field energy.

\subsubsection{Field Energy}

Turning our attention next to field energy defined in \ern{Efielddivp2} we obtain the following expression for third order terms in $\frac{1}{c}$:
\beq
E'^{[3]}_{field~12} = \int \left( \vec B'^{[0]}_1 \cdot \vec B'^{[3]}_2 +  \vec B'^{[3]}_1 \cdot \vec B'^{[0]}_2\right)   d^3 x,
\label{Efielddiv3}
\enq
in which we are reminded that there are no field contributions which are first order in $\frac{1}{c}$.
$\vec B'^{[3]}$ is calculated according to \ern{Bgeneraln} as:
\beq
\vec B'^{[3]} = \vec \nabla \times \vec A'^{[3]}.
\label{Bgeneraln3}
\enq
Taking into account \ern{Athir2} the third order correction to the magnetic field is thus:
\beq
\vec B'^{[3]} (\vec x,t) = \frac{1}{6 c^3}  \frac{d^3 I (t)}{dt^3} \oint d \tilde{\vec{l}} \times \vec \nabla R^2 (t)
= \frac{1}{3 c^3}  \frac{d^3 I (t)}{dt^3} \oint d \tilde{\vec{l}} \times \vec R(t).
\label{Bsec3}
\enq
in which we have used:
\beq
\vec \nabla R^2  = 2 R \vec \nabla R =  2 R \hat R = 2 \vec R
\enq
Hence for a static coil:
\beq
\vec B'^{[3]}_2 (\vec x,t) = 0.
\label{Bsec3b}
\enq
And thus:
\beq
E'^{[3]}_{field~12} = \int  \vec B'^{[3]}_1 \cdot \vec B'^{[0]}_2 d^3 x.
\label{Efielddiv3c}
\enq
Inserting \ern{B0} into \ern{Efielddiv0} will yield:
\beq
E'^{[3]}_{field~12} =   -\frac{1}{6 c^3}  \frac{d^3 I_1 (t)}{dt^3}  I_2   \int d^3 x  \oint d \tilde{\vec{l_1}} \times  \vec \nabla R^2_1 (t) \cdot \oint d \tilde{\vec{l_2}} \times  \vec \nabla \frac{1}{R_2 (t)}
\label{Efielddiv33}
\enq
in which we recall that $I_2$ is time independent. Using a well known identity from vector analysis we may write:
\ber
E'^{[3]}_{field~12} &=&   -\frac{1}{6 c^3}  \frac{d^3 I_1 (t)}{dt^3}    I_2   \int d^3 x  \oint \oint [
(d \tilde{\vec{l_1}} \cdot d \tilde{\vec{l_2}}) (\vec \nabla R_1^2 (t) \cdot  \vec \nabla \frac{1}{R_2 (t)})
\nonumber \\
 & - & (d \tilde{\vec{l_1}} \cdot \vec \nabla \frac{1}{R_2 (t)}) (d \tilde{\vec{l_2}} \cdot \vec \nabla R_1^2 (t)) ].
\label{Efielddiv34}
\enr
Let us look at the integral expression
\beq
int_4=\int d^3 x  \oint \oint [ (d \tilde{\vec{l_1}} \cdot \vec \nabla \frac{1}{R_2 (t)}) (d \tilde{\vec{l_2}} \cdot \vec \nabla R_1^2 (t)) ]
\label{Efielddiv35}
\enq
Such that:
\beq
E'^{[3]}_{field~12} =   -\frac{1}{6 c^3}  \frac{d^3 I_1 (t)}{dt^3}    I_2  [ \int d^3 x  \oint \oint
(d \tilde{\vec{l_1}} \cdot d \tilde{\vec{l_2}}) (\vec \nabla R_1^2 (t) \cdot  \vec \nabla \frac{1}{R_2 (t)})
- int_4].
\label{Efielddiv34b}
\enq
This is an expression of the type described in \ern{Efielddiv05d} of appendix A with $h =\frac{1}{R_2}$  and  $g = R_1^2$.
According to appendix A the expression in \ern{Efielddiv35} can be expressed as a surface integral. Assuming that
our system is contained in an infinite sphere we have according to \ern{Efielddiv05gd} and \ern{spherele}:
\beq
int_4 =\oint \oint d  l_{1n} d  l_{2m}  \lim_{r \rightarrow \infty} \oint d \Omega \ r^2 \hat r_n  \partial_m R_1^2 \frac{1}{R_2}
\label{Efielddiv05gdor3v}
\enq
Now:
\beq
\partial_m R_1^2 = 2 \vec R_{1m}
\label{parmR3}
\enq
This can be calculated  using \ern{asym1} as:
\beq
\vec R_{1m} = r (\hat r_{m}  - \frac{x_{1m}}{r}).
\label{vecRa}
\enq
Similarly according to \ern{asym3}:
\beq
\frac{1}{R}  \simeq  \frac{1}{r} \left[1 +  \frac{\hat r \cdot \vec x_2}{r}+ O \left((\frac{x_2}{r})^2\right)\right].
\label{asym3c}
\enq
Plugging \ern{vecRa} and \ern{asym3c} into \ern{Efielddiv05gdor3} we obtain:
\ber
& & \hspace{-1cm} int_4 = 2 \oint \oint d  l_{1n} d  l_{2m}  \lim_{r \rightarrow \infty} \oint d \Omega \ r^2 \hat r_n
\nonumber \\
& & \hspace{-1cm}
\left[ \hat r_{m}  - \frac{x_{1m}}{r} \right]
 \left[1 +  \frac{\hat r \cdot \vec x_2}{r}+ O \left((\frac{x_2}{r})^2\right)\right]
\label{Efielddiv05gdor3}
\enr
Now we recall that performing a loop integral over a constant $\vec C$ we obtain a null result:
\beq
\oint d  \vec l \cdot \vec C = 0
 \label{lopC3}
\enq
The only term that contribute must depend on both $\vec x_1$ and  $\vec x_2$  thus \ern{Efielddiv05gdor3} takes the form:
\beq
 int_4 = - 2 \oint \oint d  l_{1n} d  l_{2m}  \lim_{r \rightarrow \infty} \oint d \Omega \ r^2 \hat r_n
 \frac{ x_{1m}}{r}
 \left[ \frac{\hat r \cdot \vec x_2}{r}+ O \left((\frac{x_2}{r})^2\right)\right]
\label{Efielddiv05gdor3b}
\enq
And taking the limit we obtain:
\ber
 int_4 &=& - 2 \oint \oint d  l_{1n} d  l_{2m}  \oint d \Omega  \hat r_n  x_{1m}   \hat r \cdot \vec x_2
 \nonumber \\
 &=& - 2 \oint \oint d  l_{1n} d  l_{2m} x_{1m} x_{2k} \oint d \Omega  \hat r_n \hat r_k
\label{Efielddiv05gdor3c}
\enr
According to \cite{Jackson}:
\beq
\oint d \Omega  \hat r_n \hat r_k  = \frac{4 \pi}{3} \delta_{nk}
\label{rnrk}
\enq
in which $\delta_{nk}$ is a Kronecker delta, hence:
\beq
 int_4 =  - \frac{8 \pi}{3}  \oint \oint d  l_{1n} d  l_{2m} x_{1m} x_{2n} = - \frac{8 \pi}{3}  \oint \oint (d  \vec l_{1} \cdot \vec x_2)
( d  \vec l_{2} \cdot \vec x_1)
\label{Efielddiv05gdor3d}
\enq
Hence although $int_4$ is a surface integral at infinity it does not vanish which indicates radiation. The radiation contribution to the energy balance will be discussed further in the next subsection in which we shall consider the Poynting flux contribution.
Let us now look at the first part of the integral:
\beq
int_5 =    \int d^3 x  \oint \oint [(d \tilde{\vec{l_1}} \cdot d \tilde{\vec{l_2}}) (\vec \nabla R_1^2 (t) \cdot  \vec \nabla \frac{1}{R_2 (t)})].
\label{Efielddiv35g}
\enq
Such that:
\beq
E'^{[3]}_{field~12} =   -\frac{1}{6 c^3}  \frac{d^3 I_1 (t)}{dt^3}    I_2  [ int_5 - int_4].
\label{Efielddiv34c}
\enq
Now:
\beq
\vec \nabla R^2_1 (t)  \cdot  \vec \nabla \frac{1}{R_2 (t)} = \vec \nabla \cdot (R^2_1 (t)  \vec \nabla \frac{1}{R_2 (t)} )
- R^2_1 (t) \vec \nabla^2 \frac{1}{R_2 (t)}
\label{Efielddiv37}
\enq
Taking into account \ern{Efielddiv08}, we have:
\beq
\vec \nabla R^2_1 (t) \cdot  \vec \nabla \frac{1}{R_2 (t)} = \vec \nabla \cdot (R^2_1 (t) \vec \nabla \frac{1}{R_2 (t)} )
+  4 \pi R^2_1 (t) \delta (\vec R_2).
\label{Efielddiv39}
\enq
The first term in the right hand side is a divergence. Thus using Gauss theorem its volume integral will become a surface integral, the second term is a delta function. This means that there is no contribution to the volume integral from that terms unless $\vec x = \vec x_2$. We may now write \ern{Efielddiv35} as follows:
\beq
int_5 = \oint \oint (d \tilde{\vec{l_1}} \cdot d \tilde{\vec{l_2}}) [\int d a \hat n \cdot R^2_1 (t) \vec \nabla \frac{1}{R_2 (t)} +  4 \pi R^2_{12} (t) ].
\label{Efielddiv310}
\enq
Let us look at the surface integral and assume as usual that the system is contained inside an infinite sphere.
\ber
int_6 &=& \oint \oint (d \tilde{\vec{l_1}} \cdot d \tilde{\vec{l_2}}) \int d a \hat n \cdot R_1^2 (t) \vec \nabla \frac{1}{R_2 (t)}
\nonumber \\
& = &
-\oint \oint (d \tilde{\vec{l_1}} \cdot d \tilde{\vec{l_2}}) \lim_{r \rightarrow \infty} \int d \Omega \ r^2 \hat r \cdot R_1^2 (t) \frac{\vec R_2}{R_2^3 (t)}
\label{Efielddiv311}
\enr
We denote:
\beq
\vec G  = R_1^2 (t) \frac{\vec R_2}{R_2^3 (t)}
\label{Gdef}
\enq
and show in appendix \ref{asymval} that:
\beq
\oint \oint (d \tilde{\vec{l_1}} \cdot d \tilde{\vec{l_2}}) \lim_{r \rightarrow \infty} \int d \Omega \ r^2 \hat r \cdot \vec G
= \oint \oint (d \tilde{\vec{l_1}} \cdot d \tilde{\vec{l_2}})  \int d \Omega \hat r \cdot (2 (\hat r \cdot \vec x_1) \vec x_2)
\label{Efielddiv311c}
\enq
Hence:
\ber
int_6 &=& - \oint \oint (d \tilde{\vec{l_1}} \cdot d \tilde{\vec{l_2}})  \int d \Omega \hat r \cdot (2 (\hat r \cdot \vec x_1) \vec x_2)
\nonumber \\
&=&-2 \oint \oint (d \tilde{\vec{l_1}} \cdot d \tilde{\vec{l_2}})x_{1m} x_{2n}  \int d \Omega \hat r_m \hat r_n
\label{Efielddiv311d}
\enr
Now taking int account \ern{rnrk} we obtain the result:
\beq
int_6 = -\frac{8 \pi}{3} \oint \oint (d \tilde{\vec{l_1}} \cdot d \tilde{\vec{l_2}}) (\vec x_{1} \cdot \vec x_{2}) .
\label{Efielddiv311e}
\enq
Hence:
\beq
int_5 = 4 \pi \oint \oint (d \tilde{\vec{l_1}} \cdot d \tilde{\vec{l_2}})  R^2_{12} (t) -\frac{8 \pi}{3} \oint \oint (d \tilde{\vec{l_1}} \cdot d \tilde{\vec{l_2}}) (\vec x_{1} \cdot \vec x_{2}) .
\label{Efielddiv310b}
\enq
Now inserting \ern{Efielddiv310b} and \ern{Efielddiv05gdor3d} into \ern{Efielddiv34c} will yield:
\ber
E'^{[3]}_{field~12} &=&   -\frac{2 \pi}{3 c^3}  \frac{d^3 I_1 (t)}{dt^3}    I_2  [ \oint \oint (d \vec{l_1} \cdot d \vec{l_2})   R^2_{12}
- \frac{2}{3} \oint \oint (d\vec{l_1} \cdot d \vec{l_2})  (\vec x_{1} \cdot \vec x_{2})
\nonumber \\
&+& \frac{2}{3}\oint \oint (d  \vec l_{1} \cdot \vec x_2) ( d  \vec l_{2} \cdot \vec x_1)].
\label{Efielddiv319}
\enr
we will dissect the above expression into volume and surface contributions such that:
\beq
E'^{[3]}_{fieldV~12} \equiv -\frac{2 \pi}{3 c^3}  \frac{d^3 I_1 (t)}{dt^3}    I_2  \oint \oint (d \vec{l_1} \cdot d \vec{l_2})   R^2_{12}
\label{Efielddiv319V}
\enq
\ber
E'^{[3]}_{fieldS~12} &\equiv&   \frac{4 \pi}{9 c^3}  \frac{d^3 I_1 (t)}{dt^3}    I_2 \left[ \oint \oint (d\vec{l_1} \cdot d \vec{l_2}) (\vec x_{1} \cdot \vec x_{2}) \right.
\nonumber \\
&-& \left. \oint \oint (d  \vec l_{1} \cdot \vec x_2) ( d  \vec l_{2} \cdot \vec x_1)\right]
\label{Efielddiv319S}
\enr
\beq
E'^{[3]}_{field~12}= E'^{[3]}_{fieldV~12} + E'^{[3]}_{fieldS~12}
\label{Efielddiv319T}
\enq
And:
\beq
E_{fieldV~12}^{[3]} =  \frac{\mu_0}{(4 \pi)^2} E'^{[3]}_{fieldV~12}   =
-\frac{\mu_0}{24 \pi c^3}  \frac{d^3 I_1 (t)}{dt^3}    I_2  \oint \oint (d \vec{l_1} \cdot d \vec{l_2})   R^2_{12} .
\label{Efielddiv320}
\enq
It can easily seen that the change in volume energy is balance by the mechanical work done, see \ern{pow35}. For a phasor current with frequency $\omega$:
\beq
I_1 (t) = I_{1 0} e^{j \omega t}, \qquad j \equiv \sqrt{-1}
\label{phas2}
\enq
we obtain a second order correction to the mutual inductance of the form \cite{{RY}}:
\beq
M^{[3]}_{12} \equiv  \frac{j \mu_0 \omega^3}{24 \pi c^3} \oint d \vec l_1 \cdot \oint d \vec l_2 R^2_{12}
\label{CI3}
\enq
Such that:
\beq
E_{field~12}^{[3]} =  I_1 (t)  I_2  M^{[3]}_{12}.
\label{Efielddiv321}
\enq
Obviously the larger the system and the higher the frequency the more important this correction is.
We note that this term contains a $j$ indicating that this correction is resistive.
We stress that this term is not related to the relativistic engine effect and will exist even for an engine of "infinite" mass.
Of course we have unbalanced surface terms with field energy:
\ber
E_{fieldS~12}^{[3]} &=&  \frac{\mu_0}{(4 \pi)^2} E'^{[3]}_{fieldS~12}   =
\frac{\mu_0}{36 \pi c^3}  \frac{d^3 I_1 (t)}{dt^3}    I_2
\nonumber \\
 & & \hspace{-1cm} \left[ \oint \oint (d\vec{l_1} \cdot d \vec{l_2}) (\vec x_{1} \cdot \vec x_{2})
-  \oint \oint (d  \vec l_{1} \cdot \vec x_2) ( d  \vec l_{2} \cdot \vec x_1)\right] .
\label{Efielddiv322}
\enr
The only way to balance the derivative of this term in the energy equation is by a Poynting term which signifies
the generation of radiation. Finally we notice the vector identity:
\beq
(\vec x_{1} \times d\vec{l_1}) \cdot (\vec x_{2} \times d\vec{l_2}) =
(d\vec{l_1} \cdot d \vec{l_2}) (\vec x_{1} \cdot \vec x_{2})
-   (d  \vec l_{1} \cdot \vec x_2) ( d  \vec l_{2} \cdot \vec x_1)
\label{dar}
\enq
and the definition of a oriented Area:
\beq
\vec Ar \equiv \frac{1}{2}\oint \vec x \times d\vec l
\label{dar2}
\enq
and write \ern{Efielddiv322} in a compact form:
\beq
E_{fieldS~12}^{[3]}  =
\frac{\mu_0}{9 \pi c^3} I_2  \frac{d^3 I_1 (t)}{dt^3}  \vec Ar_1 \cdot \vec Ar_2.
\label{Efielddiv323}
\enq
Hence orthogonal current loops will generate a null surface field contribution.

\subsubsection{Poynting vector}

Finally we shall study the Poynting vector:
\beq
\vec S'^{[3]}_{p~12} = \vec E'^{[0]}_1 \times \vec B'^{[3]}_2 + \vec E'^{[3]}_1 \times \vec B'^{[0]}_2 +
\vec E'^{[0]}_2 \times \vec B'^{[3]}_1 + \vec E'^{[3]}_2 \times \vec B'^{[0]}_1
\label{Poyntingv3}
\enq
Taking into account that $\vec E_2^{[0]}$ is null according to \ern{E02} and $ \vec B'^{[3]}_2 $  is null according to \ern{Bsec3b}
this simplifies to:
\beq
\vec S'^{[3]}_{p~12} = \vec E'^{[3]}_1 \times \vec B'^{[0]}_2 +  \vec E'^{[3]}_2 \times \vec B'^{[0]}_1.
\label{Poyntingv3b}
\enq
Further more according to \ern{Eb3b} $\vec E'^{[3]}_2$ is also null hence:
\beq
\vec S'^{[3]}_{p~12} = \vec E'^{[3]}_1 \times \vec B'^{[0]}_2.
\label{Poyntingv3c}
\enq
The above expression can be calculated using \ern{Eb3} and \ern{B0}:
\ber
\vec S'^{[3]}_{p~12} &=& \frac{I_2}{6 c^3} \frac{d^4 I_1 (t)}{d t^4} \oint \oint d \vec l_1 \times \left( d \vec l_2 \times \frac{R_1^2 \vec R_2 }{R_2^3} \right)
\nonumber \\
&=& \frac{I_2}{6 c^3} \frac{d^4 I_1 (t)}{d t^4} \oint \oint d \vec l_1 \times \left( d \vec l_2 \times \vec G \right)
\label{Poyntingv3d}
\enr
in we have used the definition of $\vec G$ given in \ern{Gdef}. Doing some vector algebra we have:
\beq
\vec S'^{[3]}_{p~12} = \frac{I_2}{6 c^3} \frac{d^4 I_1 (t)}{d t^4} \oint \oint
\left[d \vec l_2 (d \vec l_1 \cdot \vec G) - \vec G (d \vec l_1 \cdot d \vec l_2)\right]
\label{Poyntingv3e}
\enq
Now let us calculate the Poynting flux on an infinite sphere:
\beq
\oint_S \vec S'^{[3]}_{p~12} \cdot \hat n da = \lim_{r \rightarrow \infty} \int \vec S'^{[3]}_{p~12} \cdot \hat r \  r^2 d \Omega.
\label{poyfl3}
\enq
According to \ern{Poyntingv3e} this will take the form:
\beq
\oint_S \vec S'^{[3]}_{p~12} \cdot \hat n da =\frac{I_2}{6 c^3} \frac{d^4 I_1 (t)}{d t^4} \lim_{r \rightarrow \infty} \int  \oint \oint
\left[d \vec l_2 (d \vec l_1 \cdot \vec G) - \vec G (d \vec l_1 \cdot d \vec l_2)\right] \cdot \hat r \  r^2 d \Omega.
\label{poyfl32}
\enq
According to appendix \ref{asymval} this can be calculated using the result of \ern{Efielddiv311c3}:
\ber
& &\oint_S \vec S'^{[3]}_{p~12} \cdot \hat n da =\frac{I_2}{3 c^3} \frac{d^4 I_1 (t)}{d t^4}  \int d \Omega \hat r_m  \hat r_n x_{1n}
\nonumber \\
& & \oint \oint \left[d l_{2m} (d \vec l_1 \cdot \vec x_2) - x_{2m} (d \vec l_1 \cdot d \vec l_2)\right] .
\label{poyfl33}
\enr
Applying \ern{rnrk} we obtain:
\beq
\oint_S \vec S'^{[3]}_{p~12} \cdot \hat n da =\frac{4 \pi I_2}{9 c^3} \frac{d^4 I_1 (t)}{d t^4}
 \oint \oint \left [(d \vec l_2 \cdot \vec x_1) (d \vec l_1 \cdot \vec x_2) - ( \vec x_1 \cdot \vec x_2) (d \vec l_1 \cdot d \vec l_2)\right] .
\label{poyfl34}
\enq
 Taking into account \ern{dar} and \ern{dar2} we have finally:
 \beq
\oint_S \vec S^{[3]}_{p~12} \cdot \hat n da = \frac{\mu_0}{(4 \pi)^2}  \oint_S \vec S'^{[3]}_{p~12} \cdot \hat n da=
-\frac{\mu_0}{9 \pi c^3} I_2  \frac{d^4 I_1 (t)}{dt^4}  \vec Ar_1 \cdot \vec Ar_2 .
\label{poyfl35}
\enq
 Thus the Poynting radiation flux balances that change in surface field energy given in \ern{Efielddiv323} .

\subsubsection{Intermediate account}

We conclude that the energy \ern{Econt0ord3} of the third order is indeed balanced. Mechanical work invested or extracted in the system results in increase or decrease in the field energy accordingly. To be more specific the magnetic field energy is affected by the mechanical work.
The power related to the mechanical work is according to \ern{pow35}:
\beq
Power_{12}^{[3]} =  \frac{\mu_0}{24 \pi c^3} \frac{d^4 I_1 (t)}{dt^4}  I_2 \oint \oint d \vec{l_1} \cdot d \vec{l_2} R^2_{12}
\label{pow35b}
\enq
and this is equal to minus the derivative of the volume field energy \ern{Efielddiv320}:
\beq
E_{fieldV~12}^{[3]} =  -\frac{\mu_0}{24 \pi c^3}  \frac{d^3 I_1 (t)}{dt^3}    I_2  \oint \oint (d \vec{l_1} \cdot d \vec{l_2})   R^2_{12} .
\label{Efielddiv320b}
\enq
However, for the third order in $\frac{1}{c}$ there is also a surface contribution to the field energy given in \ern{Efielddiv323}:
\beq
E_{fieldS~12}^{[3]}  =
\frac{\mu_0}{9 \pi c^3} I_2  \frac{d^3 I_1 (t)}{dt^3}  \vec Ar_1 \cdot \vec Ar_2.
\label{Efielddiv323b}
\enq
such that the total field energy is:
\beq
E^{[3]}_{field~12}= E^{[3]}_{fieldV~12} + E^{[3]}_{fieldS~12}
\label{Efielddiv319Tb}
\enq
The change in the field energy through the surface terms results in radiation as described by the Poynting flux depicted in \ern{poyfl35}:
\beq
\oint_S \vec S^{[3]}_{p~12} \cdot \hat n da = -\frac{\mu_0}{9 \pi c^3} I_2  \frac{d^4 I_1 (t)}{dt^4}  \vec Ar_1 \cdot \vec Ar_2 .
\label{poyfl35b}
\enq
curiously this flux can be avoided by configuring the loops to be orthogonal to each other \cite{RY2}.
We underline that third order contributions are not related the relativistic engine effect as non of the terms depends on the engine velocity $\vec v$, and thus the above expression will be valid even if the engine is infinitely massive and no motion occurs. We do not expect any relativistic engine contributions for orders smaller than $\frac{1}{c^4}$.
For a phasor current of frequency $\omega$ \ern{CI3} indicates a resistive relativistic correction to the classical mutual inductance which is important for large systems with high frequency.
\beq
M^{[3]}_{12} \equiv  \frac{j \mu_0 \omega^3}{24 \pi c^3} \oint d \vec l_1 \cdot \oint d \vec l_2 R^2_{12}
\label{CI3b}
\enq

\subsection{$n=4$}

Let us look at \ern{Econt0pn} and study it for the fourth order in $\frac{1}{c}$:
\beq
Power'^{[4]}_{12} = -\frac{d  E'^{[4]}_{field~12}}{dt} -\oint_S \vec S'^{[4]}_{p~12} \cdot \hat n da.
\label{Econt0ord4}
\enq

\subsubsection{Power}

We shall start by calculating $Power_{12}^{[4]}$, according to \ern{Powerpartp2lo}:
\beq
Power'^{[4]}_{12} = 4 \pi   \left( I_1 (t) \oint d \tilde{\vec{l_1}} \cdot \vec E'^{[4]}_2 (\vec x_1) + I_2 (t) \oint d \tilde{\vec{l_2}} \cdot \vec E'^{[4]}_1 (\vec x_2) \right).
\label{pow4}
\enq
The field $\vec E'^{[4]}$ can only come from the vector potential $\vec A'^{[4]}$ and $\vec A'^{[2]}$. Let us start by analyzing the contribution of the $\vec A'^{[2]}$ term, according to \ern{Asec2} this is equal to:
\beq
\vec A'^{[2]} (\vec x,t) \equiv \frac{1}{2 c^2}  \frac{d^2 I (t)}{dt^2} \oint d \tilde{\vec{l}}R(t).
\label{Asec2b}
\enq
to calculate the electric field contribution to the fourth order according to \ern{E1} we shall take the temporal derivative of \ern{Asec2b}
but keep only fourth order terms, thus we arrive at:
\beq
\vec E'^{[4]}_a (\vec x,t) \equiv \frac{1}{2 c^2}  \frac{d^2 I (t)}{dt^2} \oint d \tilde{\vec{l}} \hat R \cdot \vec v.
\label{E4a}
\enq
where we use \ern{dR0}. Hence for the static coil we have:
\beq
\vec E'^{[4]}_{2a} (\vec x,t) = 0 .
\label{E4a2}
\enq
and thus:
\beq
Power'^{[4]}_{12a} = 4 \pi  I_2 \oint d \tilde{\vec{l_2}} \cdot \vec E'^{[4]}_1 (\vec x_2).
\label{pow4a}
\enq
Plugging \ern{E4a} into \ern{pow4a}leads to the result:
\beq
Power'^{[4]}_{12a} =   \frac{2 \pi}{ c^2}  \frac{d^2 I_1 (t)}{dt^2} I_2 \oint d \tilde{\vec{l_2}} \cdot \oint d \tilde{\vec{l_1}} \hat R_{21} \cdot \vec v.
\label{pow4a2}
\enq
Using \ern{Kdef} for defining $\vec K_{122}$ we arrive at the form:
\beq
Power'^{[4]}_{12a} =   \frac{2 \pi h^2}{ c^2}  \frac{d^2 I_1 (t)}{dt^2} I_2 \vec K_{122} \cdot \vec v
\label{pow4a3}
\enq
or also:
\beq
Power_{12a}^{[4]} = \frac{\mu_0}{16 \pi^2} Power'^{[4]}_{12a} =  \frac{\mu_0 h^2}{8 \pi c^2} \frac{d^2 I_1 (t)}{dt^4}  I_2 \vec K_{122} \cdot \vec v.
\label{pow4a4}
\enq
Taking into account \ern{Pmech1b} we have:
\beq
Power_{12a}^{[4]} = \frac{d \vec P_{mech}}{d t} \cdot \vec v.
\label{pow4a5}
\enq
which is exactly the amount of mechanical power needed to drive the relativistic engine. Unfortunately more power is needed to drive the currents through the coils as will be demonstrated below.

Let us now derive an expression for $\vec A'^{[4]}$, the fourth order contribution will come from $\vec A'^{(2)}$ given in \ern{A4b2b} but also
from $\vec A'^{(4)}$ defined in \ern{A4b}:
\beq
\vec A'^{(4)} (\vec x,t) =  \frac{1}{24  c^4}  \frac{d^4}{dt^4} \left[ I (t) \oint d \tilde{\vec{l}}R^3(t) \right].
\label{Afour}
\enq
We notice that there are no contributions from $\vec A'^{(3)}$ which contains only odd powers of $\frac{1}{c}$.
It is clear that \ern{Afour} contains terms of the order of $\frac{1}{c^4}$ and higher (including terms of the order $\frac{1}{c^6}$,$\frac{1}{c^8}$ and $\frac{1}{c^{10}}$). However, it is easy to see that there are no fourth order corrections to the vector potential coming from \ern{Afour} except:
\beq
  \frac{1}{24  c^4}  \frac{d^4 I (t)}{dt^4} \oint d \tilde{\vec{l}}R^3(t).
\label{Afour2}
\enq
Taking into account \ern{Afour2} and the fourth order contributions from \ern{A4b2b} we obtain the expression:
\beq
\vec A'^{[4]} (\vec x,t) = \frac{1}{24  c^4}  \frac{d^4 I (t)}{dt^4} \oint d \tilde{\vec{l}}R^3(t)
-\frac{1}{c^2} \frac{d I(t)}{dt} \oint d \tilde{\vec{l}} \hat R \cdot \vec v
-\frac{I(t)}{2 c^2}  \oint d \tilde{\vec{l}} \hat R \cdot \frac{d \vec v}{dt}
\label{Afour3}
\enq
We shall define:
\beq
\vec A'^{[4]}_c \equiv \frac{1}{24  c^4}  \frac{d^4 I (t)}{dt^4} \oint d \tilde{\vec{l}}R^3(t)
\label{A4fourc}
\enq
for future reference. Now using  \ern{E1} and keeping only fourth order contributions we arrive at the following equation for $\vec E'^{[4]}_b$:
\ber
\vec E'^{[4]}_b &=& -\frac{1}{24  c^4}  \frac{d^5 I (t)}{dt^5} \oint d \tilde{\vec{l}}R^3(t)
+\frac{1}{c^2} \frac{d^2 I(t)}{dt^2} \oint d \tilde{\vec{l}} \hat R \cdot \vec v
\nonumber \\
&+& \frac{3}{2 c^2}  \frac{d I(t)}{dt} \oint d \tilde{\vec{l}} \hat R \cdot \frac{d \vec v}{dt}
+  \frac{I(t)}{2 c^2} \oint d \tilde{\vec{l}} \hat R \cdot \frac{d^2 \vec v}{dt^2} .
 \label{Eb4}
\enr
We shall define for future reference the mutual inductance fourth order electric field:
\beq
\vec E'^{[4]}_{mui} \equiv  -\frac{1}{24  c^4}  \frac{d^5 I (t)}{dt^5} \oint d \tilde{\vec{l}}R^3(t) = - \frac{\partial \vec A'^{[4]}_c}{\partial t}
\label{emui}
\enq
the last equation sighn is correct up to the fourth power in $\frac{1}{c}$. It is clear that for a constant current we have:
\beq
E'^{[4]}_{2b}= \frac{I_2}{2 c^2} \oint d \tilde{\vec{l_2}} \hat R_2 \cdot \frac{d^2 \vec v}{dt^2} .
\label{Eb4b}
\enq
We shall write down the total electric field correction to fourth order in $\frac{1}{c}$ for future reference:
\ber
\vec E'^{[4]} &=& \vec E'^{[4]}_a + \vec E'^{[4]}_b = -\frac{1}{24  c^4}  \frac{d^5 I (t)}{dt^5} \oint d \tilde{\vec{l}}R^3(t)
+\frac{3}{2 c^2} \frac{d^2 I(t)}{dt^2} \oint d \tilde{\vec{l}} \hat R \cdot \vec v
\nonumber \\
&+& \frac{3}{2 c^2}  \frac{d I(t)}{dt} \oint d \tilde{\vec{l}} \hat R \cdot \frac{d \vec v}{dt}
+  \frac{I(t)}{2 c^2} \oint d \tilde{\vec{l}} \hat R \cdot \frac{d^2 \vec v}{dt^2} .
 \label{ET4}
\enr

Plugging \ern{Eb4} and \ern{Eb4b} into \ern{Econt0ord4} will lead to the following expression:
\ber
Power'^{[4]}_{12b} &=& \frac{4 \pi}{c^2} \left[ \frac{1}{2} I_1 (t) I_2 \oint  \oint (d \tilde{\vec{l_1}} \cdot d \tilde{\vec{l_2}})
 \hat R_{12} \cdot \frac{d^2 \vec v}{dt^2} \right.
\nonumber \\
&-&\frac{1}{24  c^2}  \frac{d^5 I_1 (t)}{dt^5} I_2  \oint  \oint (d \tilde{\vec{l_1}} \cdot d \tilde{\vec{l_2}}) R_{21}^3(t)
\nonumber \\
&+& \frac{d^2 I_1 (t)}{dt^2}  I_2  \oint  \oint (d \tilde{\vec{l_1}} \cdot d \tilde{\vec{l_2}}) \hat R_{21} \cdot \vec v
\nonumber \\
&+&  \frac{3}{2}  \frac{d I_1(t)}{dt} I_2  \oint  \oint (d \tilde{\vec{l_1}} \cdot d \tilde{\vec{l_2}}) \hat R_{21} \cdot \frac{d \vec v}{dt}
\nonumber \\
&+& \left. \frac{1}{2} I_1 (t) I_2 \oint  \oint (d \tilde{\vec{l_1}} \cdot d \tilde{\vec{l_2}})
 \hat R_{21} \cdot \frac{d^2 \vec v}{dt^2} \right].
\label{pow4b}
\enr
Now since $\vec R_{21} = - \vec R_{12} $ the first and fifth terms cancel. And we obtain:
\ber
Power'^{[4]}_{12b} &=& \frac{4 \pi}{c^2} \left[ - \frac{1}{24  c^2}  \frac{d^5 I_1 (t)}{dt^5} I_2  \oint  \oint (d \tilde{\vec{l_1}} \cdot d \tilde{\vec{l_2}}) R_{21}^3(t) \right.
\nonumber \\
&+& \frac{d^2 I_1 (t)}{dt^2}  I_2  \oint  \oint (d\tilde{\vec{l_1}} \cdot d \tilde{\vec{l_2}}) \hat R_{21} \cdot \vec v
\nonumber \\
&+& \left. \frac{3}{2}  \frac{d I_1(t)}{dt} I_2  \oint  \oint (d\tilde{\vec{l_1}} \cdot d \tilde{\vec{l_2}}) \hat R_{21} \cdot \frac{d \vec v}{dt} \right].
\label{pow4c}
\enr
Taking into account the definition of $\vec K_{122}$ given in \ern{Kdef} we have:
\ber
Power'^{[4]}_{12b} &=& \frac{4 \pi}{c^2} \left[ - \frac{1}{24  c^2}  \frac{d^5 I_1 (t)}{dt^5} I_2  \oint  \oint (d\tilde{\vec{l_1}} \cdot d \tilde{\vec{l_2}}) R_{21}^3(t) \right.
\nonumber \\
&+& \left. \frac{d^2 I_1 (t)}{dt^2}  I_2  h^2 \vec K_{122} \cdot \vec v +  \frac{3}{2}  \frac{d I_1(t)}{dt} I_2   h^2 \vec K_{122} \cdot \frac{d \vec v}{dt} \right].
\label{pow4d}
\enr
Hence:
\ber
Power_{12b}^{[4]} &=& \frac{\mu_0}{16 \pi^2} Power'^{[4]}_{12b} =
\left[ - \frac{\mu_0}{96 \pi  c^4}  \frac{d^5 I_1 (t)}{dt^5} I_2  \oint  \oint (d\tilde{\vec{l_1}} \cdot d \tilde{\vec{l_2}}) R_{21}^3(t) \right.
\nonumber \\
& & \hspace{-2cm} +\left. \frac{\mu_0}{4 \pi  c^2}  \frac{d^2 I_1 (t)}{dt^2}  I_2  h^2 \vec K_{122} \cdot \vec v +
\frac{3 \mu_0}{8 \pi  c^2}    \frac{d I_1(t)}{dt} I_2   h^2 \vec K_{122} \cdot \frac{d \vec v}{dt} \right].
\label{pow45}
\enr
Now using the expression for the relativistic engine mechanical momentum $\vec P_{mech}$ given in \ern{Pmech1b} we have:
\ber
Power_{12b}^{[4]} &=& \frac{\mu_0}{16 \pi^2} Power'^{[4]}_{12b} =
 - \frac{\mu_0}{96 \pi  c^4}  \frac{d^5 I_1 (t)}{dt^5} I_2  \oint  \oint (d\tilde{\vec{l_1}} \cdot d \tilde{\vec{l_2}}) R_{21}^3(t)
\nonumber \\
& & +2 \frac{d \vec P_{mech}}{d t} \cdot \vec v + 3 \vec P_{mech}\cdot \frac{d \vec v}{dt}
\nonumber \\
&=& - \frac{\mu_0}{96 \pi  c^4}  \frac{d^5 I_1 (t)}{dt^5} I_2  \oint  \oint d (\vec{l_1} \cdot d \vec{l_2}) R_{21}^3 +5 \vec P_{mech}\cdot \frac{d \vec v}{dt}
\label{pow46}
\enr
Hence the total mechanical work done in the fourth order can be calculated using \ern{pow4a5} and \ern{pow46} as:
\ber
Power_{12}^{[4]}&=&Power_{12a}^{[4]}+Power_{12b}^{[4]}
\nonumber \\
&=& - \frac{\mu_0}{96 \pi  c^2}  \frac{d^5 I_1 (t)}{dt^5} I_2  \oint  \oint (d\vec{l_1} \cdot d \vec{l_2}) R_{21}^3 +6 \vec P_{mech}\cdot \frac{d \vec v}{dt}
\nonumber \\
&=& - \frac{\mu_0}{96 \pi  c^2}  \frac{d^5 I_1 (t)}{dt^5} I_2  \oint  \oint (d\vec{l_1} \cdot d \vec{l_2}) R_{21}^3 +6 \frac{d E_{mech}}{dt}.
\label{pow47}
\enr
We notice that power invested is mechanical work to fourth order in $\frac{1}{c}$ has two parts. One which clearly is not related the relativistic engine effect and the other which clearly is. As related to the mechanical power needed to operate the relativistic engine it is six times greater then the change in kinetic energy of the engine itself, the rest of the power is invested in driving the currents through the coils. An additional part which is related to the fifth derivative of the current is not connected to relativistic engine effect and will exist even for an infinitely heavy engine.

\subsubsection{Field Energy}

Turning our attention next to field energy defined in \ern{Efielddivp2} we obtain the following expression for fourth order term in $\frac{1}{c}$:
\ber
E'^{[4]}_{field~12} &=& \int \left( \frac{1}{c^2} \left( \vec E'^{[0]}_1 \cdot \vec E'^{[2]}_2 + \vec E'^{[2]}_1 \cdot \vec E'^{[0]}_2 \right)\right.
\nonumber \\
&+&
\left. \vec B'^{[0]}_1 \cdot \vec B'^{[4]}_2 + \vec B'^{[2]}_1 \cdot \vec B'^{[2]}_2 + \vec B'^{[4]}_1 \cdot \vec B'^{[0]}_2\right)   d^3 x
\label{Efielddiv4}
\enr
in which we are reminded that there are no field contributions which are first order in $\frac{1}{c}$.
According to \ern{E02} $\vec E'^{[0]}_2 = 0$, and according to \ern{Bsec2b} $\vec B'^{[2]}_2  = 0$, hence the above equation simplifies as follows:
\beq
E'^{[4]}_{field~12} =\int \left( \frac{1}{c^2} \vec E'^{[0]}_1 \cdot \vec E'^{[2]}_2   + \vec B'^{[0]}_1 \cdot \vec B'^{[4]}_2 +  \vec B'^{[4]}_1 \cdot \vec B'^{[0]}_2\right)   d^3 x.
\label{Efielddiv42}
\enq
The field energy can be clearly partitioned to electric field  and magnetic field contributions:
\ber
E'^{[4]}_{field~12} &=& E'^{[4]}_{Efield~12} + E'^{[4]}_{Mfield~12}
\nonumber \\
 E'^{[4]}_{Efield~12} &=& \frac{1}{c^2} \int  \ \vec E'^{[0]}_1 \cdot \vec E'^{[2]}_2
 \nonumber \\
 E'^{[4]}_{Mfield~12} &=& \int \left( \vec B'^{[0]}_1 \cdot \vec B'^{[4]}_2 +  \vec B'^{[4]}_1 \cdot \vec B'^{[0]}_2\right)   d^3 x.
\label{Efielddiv43}
\enr
We begin with evaluating $E'^{[4]}_{Efield~12}$ by using $\vec E'^{[0]}_1$ of \ern{E0} and $\vec E'^{[2]}_2$ of \ern{E22}.
Obtaining:
\beq
 E'^{[4]}_{Efield~12} = \frac{I_2}{c^2} \frac{d I_1(t)}{dt} \oint  \oint (d \vec{l_1} \cdot d \vec{l_2}) \vec v \cdot
 \int d^3 x  \frac{\vec R_2}{R_2^2 R_1}
\label{Efielddiv44}
\enq
We show in appendix \ref{Qfun} (see also \cite{AY2}) that:
\beq
2\pi \hat R_{12} =  \int d^3 x  \frac{\vec R_2}{R_2^2 R_1},
\enq
hence:
\beq
 E'^{[4]}_{Efield~12} = \frac{2 \pi I_2}{c^2} \frac{d I_1(t)}{dt}\vec v \cdot \oint  \oint (d \vec{l_1} \cdot d \vec{l_2})   \hat R_{12}.
\label{Efielddiv45}
\enq
Taking into account \ern{Kdef} we obtain:
\beq
 E'^{[4]}_{Efield~12} = -\frac{2\pi I_2}{c^2} \frac{d I_1(t)}{dt}\vec v \cdot \vec K_{122}.
\label{Efielddiv46}
\enq
hence:
\beq
E_{Efield~12}^{[4]} =  \frac{\mu_0}{(4 \pi)^2} E'^{[4]}_{Efield~12}   =
-  \frac{\mu_0}{8 \pi c^2}  \frac{d I_1 (t)}{dt}   I_2   \vec v \cdot \vec K_{122}.
\label{Efielddiv47}
\enq
using the mechanical momentum  \ern{Pmech1b} we thus obtain:
\beq
E_{Efield~12}^{[4]} =  -   \vec v \cdot\vec P_{mech} = -2 E_{mech}.
\label{Efielddiv48}
\enq
Turning our attention to the magnetic part of the field energy we notice that a fourth order correction of the magnetic field $\vec B'^{[4]}$ is
needed, this can be calculated according to \ern{Bgeneraln} as:
\beq
\vec B'^{[4]} = \vec \nabla \times \vec A'^{[4]}.
\label{Bgeneraln4}
\enq
Taking into account \ern{Afour3} the fourth order correction to the magnetic field is thus:
\ber
\vec B'^{[4]} (\vec x,t) &=& -\frac{I(t)}{2 c^2}  \oint \vec \nabla (\hat R \cdot \frac{d \vec v}{dt}) \times d \tilde{\vec{l}}
-\frac{1}{c^2} \frac{d I(t)}{dt} \oint \vec \nabla (\hat R \cdot \vec v) \times d \tilde{\vec{l}}
\nonumber \\
&+& \frac{1}{24  c^4}  \frac{d^4 I (t)}{dt^4} \oint \vec \nabla R^3(t) \times d \tilde{\vec{l}} .
\label{Bsec4}
\enr
Hence for a static current:
\beq
\vec B'^{[4]}_2 (\vec x,t) = -\frac{I_2}{2 c^2}  \oint \vec \nabla (\hat R_2 \cdot \frac{d \vec v}{dt}) \times d \tilde{\vec{l_2}}.
\label{Bsec4b}
\enq
We shall find it convenient to label the different terms of the magnetic field of the fourth order:
\ber
\vec B'^{[4]} (\vec x,t) &=& \vec B'^{[4]}_a +  \vec B'^{[4]}_b +  \vec B'^{[4]}_c
\nonumber \\
\vec B'^{[4]}_a &=& -\frac{I(t)}{2 c^2}  \oint \vec \nabla (\hat R \cdot \frac{d \vec v}{dt}) \times d \tilde{\vec{l}}
\nonumber \\
\vec B'^{[4]}_b &=& -\frac{1}{c^2} \frac{d I(t)}{dt} \oint \vec \nabla (\hat R \cdot \vec v) \times d \tilde{\vec{l}}
\nonumber \\
\vec B'^{[4]}_c &=& \frac{1}{24  c^4}  \frac{d^4 I (t)}{dt^4} \oint \vec \nabla R^3(t) \times d \tilde{\vec{l}} .
\label{Bsec4c}
\enr
And thus the magnetic energy can also be partitioned:
\ber
 E'^{[4]}_{Mfield~12} &=&  E'^{[4]}_{Mfield~120}+E'^{[4]}_{Mfield~12a}+E'^{[4]}_{Mfield~12b}+E'^{[4]}_{Mfield~12c}
 \nonumber \\
 E'^{[4]}_{Mfield~120} &=&  \int  \vec B'^{[0]}_1 \cdot \vec B'^{[4]}_2  d^3 x
 \nonumber \\
 E'^{[4]}_{Mfield~12a} &=&  \int  \vec B'^{[4]}_{1a} \cdot \vec B'^{[0]}_2  d^3 x
 \nonumber \\
 E'^{[4]}_{Mfield~12b} &=&  \int  \vec B'^{[4]}_{1b} \cdot \vec B'^{[0]}_2  d^3 x
 \nonumber \\
 E'^{[4]}_{Mfield~12c} &=&  \int  \vec B'^{[4]}_{1c} \cdot \vec B'^{[0]}_2  d^3 x.
\label{Efielddiv4c}
\enr
We shall start by evaluating $ E'^{[4]}_{Mfield~120}$. Using \ern{B0} and \ern{Bsec4b} we obtain:
\beq
E'^{[4]}_{Mfield~120} =   -\frac{I_1 (t)  I_2 }{2 c^2}     \int d^3 x \oint \oint
\left(d \tilde{\vec{l_2}} \times  \vec \nabla (\hat R_2 (t) \cdot \frac{d \vec v}{dt}) \right)\cdot
\left(d \tilde{\vec{l_1}} \times  \vec \nabla \frac{1}{R_1 (t)}\right)
\label{EnM0}
\enq
Using a well known identity from vector analysis we may write:
\ber
E'^{[4]}_{Mfield~120} &=&   -\frac{I_1 (t)  I_2 }{2 c^2}    \int d^3 x  \oint \oint [
\left(d \tilde{\vec{l_1}} \cdot d \tilde{\vec{l_2}}\right) \left(\vec \nabla (\hat R_2 (t) \cdot \frac{d \vec v}{dt}) \cdot  \vec \nabla \frac{1}{R_1 (t)}\right)
\nonumber \\
 & - & \left(d \tilde{\vec{l_2}} \cdot \vec \nabla \frac{1}{R_1 (t)}\right) \left(d \tilde{\vec{l_1}} \cdot \vec \nabla (\hat R_2 (t) \cdot \frac{d \vec v}{dt})\right) ].
\label{EnM02}
\enr
Let us look at the integral expression
\beq
int_7=\int d^3 x  \oint \oint [ (d \tilde{\vec{l_2}} \cdot \vec \nabla \frac{1}{R_1 (t)}) (d \tilde{\vec{l_1}} \cdot \vec \nabla (\hat R_2 (t) \cdot \frac{d \vec v}{dt})) ]
\label{int745}
\enq
This is an expression of the type described in \ern{Efielddiv05d} of appendix A with $g =\frac{1}{R_1}$  and  $h = \hat R_2 (t) \cdot \frac{d \vec v}{dt}$.
According to appendix A the expression in \ern{int745} can be expressed as a surface integral. Assuming that
our system is contained in an infinite sphere we have according to \ern{Efielddiv05gd} and \ern{spherele}:
\beq
int_7 =\oint \oint d  l_{1n} d  l_{2m}  \lim_{r \rightarrow \infty} \oint d \Omega \ r^2 \hat r_n  \partial_m \frac{1}{R_1} \hat R_2  \cdot \frac{d \vec v}{dt}
\label{Efielddiv05gdor4v}
\enq
The following asymptotic expressions will now come in handy (see \ern{hatRa} and \ern{Efielddiv212}):
\beq
\hat R_2 = \hat r+ \frac{1}{r}(\hat r (\hat r \cdot \vec x_2) - \vec x_{2}) + O \left((\frac{x_2}{r})^2\right).
\label{hatRasym}
\enq
\beq
\partial_m \frac{1}{R_1} = -\frac{1}{r^2} \left[\hat r_m + \frac{1}{r}(3 (\hat r  \cdot \vec x_1) \hat r_m - \vec x_{1m})+ O \left((\frac{x_1}{r})^2\right) \right].
\label{parm1oR1}
\enq
Inserting \ern{hatRasym} and \ern{parm1oR1} into \ern{Efielddiv05gdor4v} and taking the limit will yield:
\beq
int_7 =-\oint \oint d  l_{1n} d  l_{2m}  \oint d \Omega \ \hat r_n  \hat r_m \hat r  \cdot \frac{d \vec v}{dt}
\label{Efielddiv05gdor4vb}
\enq
However, according to \ern{lopC3} a closed loop integral over a constant is null hence:
\beq
int_7 = 0
\label{Efielddiv05gdor4vc}
\enq
And \ern{EnM02} simplifies to:
\beq
E'^{[4]}_{Mfield~120} =  -\frac{I_1 (t)  I_2 }{2 c^2}    \int d^3 x  \oint \oint [
\left(d \tilde{\vec{l_1}} \cdot d \tilde{\vec{l_2}}\right) \left(\vec \nabla (\hat R_2 (t) \cdot \frac{d \vec v}{dt}) \cdot  \vec \nabla \frac{1}{R_1 (t)}\right)].
\label{EnM03}
\enq
Now:
\beq
\vec \nabla (\hat R_2  \cdot \frac{d \vec v}{dt})  \cdot  \vec \nabla \frac{1}{R_1 } = \vec \nabla \cdot \left((\hat R_2  \cdot \frac{d \vec v}{dt})  \vec \nabla \frac{1}{R_1 }\right ) - (\hat R_2  \cdot \frac{d \vec v}{dt}) \vec \nabla^2 \frac{1}{R_1 }
\label{Efielddiv47z}
\enq
Taking into account \ern{Efielddiv08}, we have:
\beq
\vec \nabla (\hat R_2  \cdot \frac{d \vec v}{dt})  \cdot  \vec \nabla \frac{1}{R_1 } = \vec \nabla \cdot \left((\hat R_2  \cdot \frac{d \vec v}{dt})  \vec \nabla \frac{1}{R_1 }\right ) +  4 \pi (\hat R_2  \cdot \frac{d \vec v}{dt}) \delta (\vec R_1).
\label{Efield48}
\enq
Plugging \ern{Efield48} into \ern{EnM03} and using Gauss theorem we obtain:
\beq
E'^{[4]}_{Mfield~120} =  -\frac{I_1 (t)  I_2 }{2 c^2}  \oint \oint \left(d \vec{l_1} \cdot d \vec{l_2}\right)  [\int da \hat n \cdot  (\hat R_2  \cdot \frac{d \vec v}{dt})  \vec \nabla \frac{1}{R_1 } + 4 \pi \hat R_{12}  \cdot \frac{d \vec v}{dt}].
\label{EnM04}
\enq
Let us perform the surface integral on an infinite sphere as usual and look at the integral:
\beq
int_8 =\oint \oint \left(d \vec{l_1} \cdot d \vec{l_2}\right) \lim_{r \rightarrow \infty} \oint d \Omega \ r^2 \hat r  \cdot  (\hat R_2  \cdot \frac{d \vec v}{dt})  \vec \nabla \frac{1}{R_1}
\label{Efielddiv05gd4v}
\enq
Now \ern{parm1oR1} takes the asymptotic form:
\beq
\vec \nabla \frac{1}{R_1} = -\frac{1}{r^2} \left[\hat r + \frac{1}{r}(3 (\hat r  \cdot \vec x_1) \hat r - \vec x_{1})+ O \left((\frac{x_1}{r})^2\right) \right].
\label{parm1oR1b}
\enq
Using \ern{hatRasym} and \ern{parm1oR1b} in \ern{Efielddiv05gd4v} and taking the limit:
\beq
int_8 = - \oint \oint \left(d \vec{l_1} \cdot d \vec{l_2}\right) \oint d \Omega \  \hat r  \cdot  (\hat r  \cdot \frac{d \vec v}{dt})  \hat r
= - \oint \oint \left(d \vec{l_1} \cdot d \vec{l_2}\right) \oint d \Omega \  (\hat r  \cdot \frac{d \vec v}{dt})
\label{Efielddiv05gd4vb}
\enq
However, according to \ern{lopC3} a closed loop integral over a constant is null hence:
\beq
int_8 = 0
\label{Efielddiv05gd4vc}
\enq
and thus \ern{EnM04} simplifies
\beq
E'^{[4]}_{Mfield~120} =  -\frac{2 \pi I_1 (t)  I_2 }{c^2}  \oint \oint \left(d \vec{l_1} \cdot d \vec{l_2}\right)  \hat R_{12}  \cdot \frac{d \vec v}{dt}.
\label{EnM05}
\enq
Taking into account the definition \ern{Kdef} this is simplified to the form:
\beq
E'^{[4]}_{Mfield~120} =  2 \pi I_1 (t)  I_2 \frac{ h^2 }{c^2}  \vec K_{122}\cdot \frac{d \vec v}{dt}.
\label{EnM06}
\enq
Next we turn our attention to $E'^{[4]}_{Mfield~12a}$ (see \ern{Efielddiv4c}), using \ern{B0} and \ern{Bsec4c} we obtain:
\beq
E'^{[4]}_{Mfield~12a} =   -\frac{I_1 (t)  I_2 }{2 c^2}     \int d^3 x \oint \oint
\left(d \tilde{\vec{l_1}} \times  \vec \nabla (\hat R_1 (t) \cdot \frac{d \vec v}{dt}) \right)\cdot
\left(d \tilde{\vec{l_2}} \times  \vec \nabla \frac{1}{R_2 (t)}\right).
\label{EnMa}
\enq
However, this integral is the same as the integral given in \ern{EnM0} with the indices $1$ and $2$ interchanged. It immediately follows that
$E'^{[4]}_{Mfield~12a}$ is equal to right hand side of \ern{EnM06} with the indices $1$ and $2$ interchanged, thus:
\beq
E'^{[4]}_{Mfield~12a} =  2 \pi I_1 (t)  I_2 \frac{ h^2 }{c^2}  \vec K_{212}\cdot \frac{d \vec v}{dt}.
\label{EnMa2}
\enq
However, according to \ern{Kdef}:
\beq
\vec K_{212} = -\vec K_{122}.
\label{Kdef2}
\enq
Thus we obtain:
\beq
E'^{[4]}_{Mfield~12a} =  - E'^{[4]}_{Mfield~120}.
\label{EnMa3}
\enq
And \ern{Efielddiv4c} simplifies to the form:
\beq
 E'^{[4]}_{Mfield~12} = E'^{[4]}_{Mfield~12b}+E'^{[4]}_{Mfield~12c}
\label{EnMt3}
\enq
this is to be expected as the energy terms should not depend on acceleration but only on velocity. We now turn our attention
to $E'^{[4]}_{Mfield~12b}$ defined in \ern{Efielddiv4c}. Using \ern{B0} and \ern{Bsec4c} we obtain:
\beq
E'^{[4]}_{Mfield~12b} =   -\frac{  I_2 }{c^2}   \frac{d I_1 (t)}{dt}  \int d^3 x \oint \oint
\left(d \tilde{\vec{l_1}} \times  \vec \nabla (\hat R_1 (t) \cdot \vec v ) \right)\cdot
\left(d \tilde{\vec{l_2}} \times  \vec \nabla \frac{1}{R_2 (t)}\right).
\label{EnMb}
\enq
Using a well known identity from vector analysis we may write:
\ber
E'^{[4]}_{Mfield~12b} &=&   -\frac{  I_2 }{c^2}   \frac{d I_1 (t)}{dt} \int d^3 x  \oint \oint [
\left(d \vec{l_1} \cdot d \vec{l_2}\right) \left(\vec \nabla (\hat R_1  \cdot \vec v )) \cdot  \vec \nabla \frac{1}{R_2 }\right)
\nonumber \\
 & - & \left(d \vec{l_1} \cdot \vec \nabla \frac{1}{R_2 }\right) \left(d \vec{l_2} \cdot \vec \nabla (\hat R_1  \cdot \vec v )\right) ].
\label{EnMb2}
\enr
Let us look at the integral expression
\beq
int_9=\int d^3 x  \oint \oint \left(d \vec{l_1} \cdot \vec \nabla \frac{1}{R_2 }\right) \left(d \vec{l_2} \cdot \vec \nabla (\hat R_1  \cdot \vec v )\right)
\label{Efielddivb45}
\enq
This is an expression of the type described in \ern{Efielddiv05d} of appendix A with $g = \hat R_1  \cdot \vec v $  and  $h = \frac{1}{R_2 }$.
According to appendix A the expression in \ern{Efielddiv45} can be expressed as a surface integral. Assuming that
our system is contained in an infinite sphere we have according to \ern{Efielddiv05gd} and \ern{spherele}:
\beq
int_9 =\oint \oint d  l_{1n} d  l_{2m}  \lim_{r \rightarrow \infty} \int d \Omega \ r^2 \hat r_n  \partial_m (\hat R_1  \cdot \vec v ) \frac{1}{R_2 }
\label{Efielddiv05b4v}
\enq
Now:
\beq
\vec \nabla (\hat R  \cdot \vec v) = v_i \vec \nabla (\frac{R_i}{R})=  v_i (\frac{\hat x_i}{R} - \frac{R_i \hat R}{R^2} )
= \frac{1}{R} (\vec v - (\vec v \cdot \hat R) \hat R)
\label{Efielddiv05b4v1}
\enq
Taking into account the asymptotic expressions in \ern{hatRasym} and \ern{asym3} the limit of \ern{Efielddiv05b4v} takes the following form:
\beq
int_9 =\oint \oint d  l_{1n} d  l_{2m}  \int d \Omega \ \hat r_n  ( v_m - (\vec v \cdot \hat r) \hat r_m)
\label{Efielddiv05b4v2}
\enq
However, according to \ern{lopC3} a closed loop integral over a constant is null hence:
\beq
int_9 = 0
\label{Efielddiv05b4v3}
\enq
And \ern{EnMb2} simplifies to:
\beq
E'^{[4]}_{Mfield~12b} =  -\frac{  I_2 }{c^2}   \frac{d I_1 (t)}{dt} \int d^3 x  \oint \oint
\left(d \vec{l_1} \cdot d \vec{l_2}\right) \left(\vec \nabla (\hat R_1  \cdot \vec v ) \cdot  \vec \nabla \frac{1}{R_2 }\right).
\label{EnMb3}
\enq
Now:
\beq
\vec \nabla (\hat R_1  \cdot \vec v)  \cdot  \vec \nabla \frac{1}{R_2 } = \vec \nabla \cdot \left((\hat R_1  \cdot \vec v)  \vec \nabla \frac{1}{R_2 }\right ) - (\hat R_1  \cdot \vec v ) \vec \nabla^2 \frac{1}{R_2 }
\label{Efielddiv4b7}
\enq
Taking into account \ern{Efielddiv08}, we have:
\beq
\vec \nabla (\hat R_1  \cdot \vec v)  \cdot  \vec \nabla \frac{1}{R_2 } = \vec \nabla \cdot \left((\hat R_1  \cdot \vec v)  \vec \nabla \frac{1}{R_2 }\right ) +  4 \pi (\hat R_1  \cdot \vec v ) \delta (\vec R_2)
\label{Efielddiv4bb7}
\enq
Plugging \ern{Efielddiv4b7} into \ern{EnMb3} and using Gauss theorem we obtain:
\beq
E'^{[4]}_{Mfield~12b} =   -\frac{  I_2 }{c^2}   \frac{d I_1 (t)}{dt} \oint \oint \left(d \vec{l_1} \cdot d \vec{l_2}\right)  [\int da \hat n \cdot  (\hat R_1  \cdot \vec v) \vec \nabla \frac{1}{R_2 } + 4 \pi \hat R_{21}  \cdot \vec v].
\label{EnMb4}
\enq
Let us perform the surface integral on an infinite sphere as usual and look at the integral:
\beq
int_{10} =\oint \oint \left(d \vec{l_1} \cdot d \vec{l_2}\right) \lim_{r \rightarrow \infty} \oint d \Omega \ r^2 \hat r  \cdot  (\hat R_1  \cdot \vec v) \vec \nabla \frac{1}{R_2 }
\label{int10v}
\enq
Using \ern{hatRasym} and \ern{parm1oR1b} in \ern{int10v} and taking the limit we obtain:
\beq
int_{10} = - \oint \oint \left(d \vec{l_1} \cdot d \vec{l_2}\right) \oint d \Omega \  \hat r  \cdot  (\hat r  \cdot \vec v)  \hat r
= - \oint \oint \left(d \vec{l_1} \cdot d \vec{l_2}\right) \oint d \Omega \  (\hat r  \cdot \vec v)
\label{Efielddiv05gd4vbb}
\enq
However, according to \ern{lopC3} a closed loop integral over a constant is null hence:
\beq
int_{10} = 0
\label{Efielddiv05b4vc}
\enq
and thus \ern{EnMb4} simplifies to:
\beq
E'^{[4]}_{Mfield~12b} =  4 \pi \frac{  I_2 }{c^2}   \frac{d I_1 (t)}{dt}  \oint \oint \left(d \vec{l_1} \cdot d \vec{l_2}\right)  \hat R_{12}  \cdot \vec v.
\label{EnMb5}
\enq
Taking into account the definition in \ern{Kdef} this is simplified to the form:
\beq
E'^{[4]}_{Mfield~12b} =   -4 \pi I_2  \frac{d I_1 (t)}{dt} \frac{ h^2 }{c^2}  \vec K_{122}\cdot \vec v.
\label{EnMb6}
\enq
Hence:
\beq
E^{[4]}_{Mfield~12b} =  \frac{\mu_0}{(4 \pi)^2} E'^{[4]}_{Mfield~12b}   =
 - \frac{\mu_0}{4 \pi} I_2  \frac{d I_1 (t)}{dt} \frac{ h^2 }{c^2}  \vec K_{122}\cdot \vec v.
\label{EnMb7}
\enq
Taking into account the mechanical momentum \ern{Pmech1b} and the mechanical energy  \ern{Emech} this can be written as:
\beq
E^{[4]}_{Mfield~12b} =  - 2 \vec P_{mech} \cdot \vec v = - 4 E_{mech}.
\label{EnMb8}
\enq
Finally we turn we turn our attention to $E'^{[4]}_{Mfield~12c}$ defined in \ern{Efielddiv4c}, by taking the time derivative of $E'^{[4]}_{Mfield~12c}$ and keeping only terms of the fourth order in $\frac{1}{c}$:
\beq
\frac{d E'^{[4]}_{Mfield~12c}}{d t} =  \int  \frac{\partial \vec B'^{[4]}_{1c}}{\partial t} \cdot \vec B'^{[0]}_2  d^3 x
\label{EnMc1}
\enq
According to \ern{Bgeneraln4}, \ern{A4fourc}  and \ern{emui} this can be written as:
\beq
\frac{d E'^{[4]}_{Mfield~12c}}{d t} = - \int  \vec \nabla \times \vec E'^{[4]}_{1~mui} \cdot \vec B'^{[0]}_2  d^3 x
\label{EnMc2}
\enq
However, according to a well known vector analysis identity:
\beq
\vec \nabla \times \vec E'^{[4]}_{1~mui} \cdot \vec B'^{[0]}_2  = \vec \nabla \cdot (\vec E'^{[4]}_{1~mui} \times \vec B'^{[0]}_2)
+ \vec E'^{[4]}_{1~mui} \cdot \vec \nabla \times \vec B'^{[0]}_2
\label{EnMc2b}
\enq
Now to zeroth order in $\frac{1}{c}$ Maxwell equations dictate that:
\beq
\vec \nabla \times \vec B'^{[0]}_2 = 4 \pi \vec J_2
\label{EnMc2c}
\enq
Thus we may write:
\beq
\vec \nabla \times \vec E'^{[4]}_{1~mui} \cdot \vec B'^{[0]}_2  = \vec \nabla \cdot (\vec E'^{[4]}_{1~mui} \times \vec B'^{[0]}_2)
+ 4 \pi\vec E'^{[4]}_{1~mui} \cdot \vec J_2
\label{EnMc2d}
\enq
Plugging \ern{EnMc2d} into \ern{EnMc2} and using Gauss theorem we obtain:
 \beq
\frac{d E'^{[4]}_{Mfield~12c}}{d t} = - \int da \ \hat n \cdot (\vec E'^{[4]}_{1~mui} \times \vec B'^{[0]}_2)
 -  4 \pi I_2  \oint d \vec l_2  \cdot \vec E'^{[4]}_{1~mui} (\vec x_2)
\label{EnMc3}
\enq
Now taking into account \ern{emui} this can be written  as:
\ber
\frac{d E'^{[4]}_{Mfield~12c}}{d t} &=& \frac{1}{24  c^4}  \frac{d^5 I_1 (t)}{dt^5}
[ \int da \ \hat n \cdot (\oint d \tilde{\vec{l_1}}R_1^3(t)\times \vec B'^{[0]}_2)
 \nonumber \\
 &+&  4 \pi I_2  \oint \oint (d \vec l_1  \cdot d \vec l_2)  R^3_{12}]
\label{EnMc4}
\enr
to the fourth order in $\frac{1}{c}$ we may write:
\ber
\frac{d E'^{[4]}_{Mfield~12c}}{d t} &=& \frac{d}{dt} \left\{\frac{1}{24  c^4}  \frac{d^4 I_1 (t)}{dt^4}
[ \int da \ \hat n \cdot (\oint d \tilde{\vec{l_1}}R_1^3(t)\times \vec B'^{[0]}_2) \right.
 \nonumber \\
 &+&  \left. 4 \pi I_2  \oint \oint (d \vec l_1  \cdot d \vec l_2)  R^3_{12}] \right\}
\label{EnMc5}
\enr
Hence up to a constant:
\ber
E'^{[4]}_{Mfield~12c} &=& \frac{1}{24  c^4}  \frac{d^4 I_1 (t)}{dt^4}
[ \int da \ \hat n \cdot (\oint d \tilde{\vec{l_1}}R_1^3(t)\times \vec B'^{[0]}_2)
 \nonumber \\
 &+&  4 \pi I_2  \oint \oint (d \vec l_1  \cdot d \vec l_2)  R^3_{12}]
\label{EnMc6}
\enr
We notice that this magnetic energy term has a surface and volume contributions as follows:
\beq
E'^{[4]}_{Mfield~12c}= E'^{[4]}_{MfieldV~12c} + E'^{[4]}_{MfieldS~12c}
\label{EMcT}
\enq
In which:
\beq
E_{MfieldV~12c}^{[4]} =  \frac{\mu_0}{(4 \pi)^2} E'^{[4]}_{MfieldV~12c}   =
\frac{\mu_0}{96 \pi c^4}  \frac{d^4 I_1 (t)}{dt^4}    I_2  \oint \oint (d \vec{l_1} \cdot d \vec{l_2})   R^3_{12}.
\label{EMcT1}
\enq
It can easily seen that the change in volume energy is balance by the mechanical work done (see \ern{pow47}). For a phasor current with frequency $\omega$ defined in \ern{phas2} we obtain a fourth order correction to the mutual inductance of the form
\beq
M^{[4]}_{12} \equiv  \frac{\mu_0 \omega^4}{96 \pi c^4} \oint d \vec l_1 \cdot \oint d \vec l_2 R^3_{12}
\label{CI4}
\enq
Such that:
\beq
E_{MfieldV~12c}^{[4]} =  I_1 (t)  I_2  M^{[4]}_{12}.
\label{Efielddi421}
\enq
Obviously the larger the system and the higher the frequency the more important this correction is.
We stress that this term is not related to the relativistic engine effect and will exist even for an engine of "infinite" mass.
The surface terms of the field energy are:
\ber
E_{MfieldS~12c}^{[4]} &=&  \frac{\mu_0}{(4 \pi)^2} E'^{[4]}_{MfieldS~12c}
 \nonumber \\
& & \hspace{-2cm} =\frac{\mu_0}{384 \pi^2 c^4}  \frac{d^4 I_1 (t)}{dt^4}    I_2
\int da \ \hat n \cdot (\oint d \tilde{\vec{l_1}}R_1^3(t)\times \vec B'^{[0]}_2)
\label{Efielddi422}
\enr
We recall that the derivative of this term is:
\beq
\frac{d E'^{[4]}_{MfieldS~12c}}{d t} = - \int da \ \hat n \cdot (\vec E'^{[4]}_{1~mui} \times \vec B'^{[0]}_2)
\label{Efielddi422b}
\enq
This term is not balanced by mechanical work and thus the only way to balance the derivative of this term in the energy equation is by a Poynting term which signifies the generation of radiation and will be discussed in the next section. We notice that this term will not exist if $\frac{d^4 I_1 (t)}{dt^4}=0$ but the relativistic engine effect will still exist provided there is a second order derivative to the current.

The total magnetic energy can calculated by plugging \ern{EMcT} and \ern{EnMb8} into \ern{EnMt3}. This will partitioned into a volume and surface terms as follows:
\beq
E^{[4]}_{Mfield~12}= E^{[4]}_{MfieldV~12} + E^{[4]}_{MfieldS~12}
\label{EMT}
\enq
in which:
\beq
E^{[4]}_{MfieldV~12} = - 4 E_{mech} + \frac{\mu_0}{96 \pi c^4}  \frac{d^4 I_1 (t)}{dt^4}    I_2  \oint \oint (d \vec{l_1} \cdot d \vec{l_2})   R^3_{12}.
\label{EMVT}
\enq
and:
\beq
 E^{[4]}_{MfieldS~12} =  E^{[4]}_{MfieldS~12c}
\label{EMST}
\enq
Finally we may calculate the total field energy by plugging \ern{EMT} and \ern{Efielddiv48} into \ern{Efielddiv43}. The total field energy will partitioned into a volume and surface terms as follows:
\beq
E^{[4]}_{field~12}= E^{[4]}_{fieldV~12} + E^{[4]}_{fieldS~12}
\label{ET}
\enq
in which:
\beq
E^{[4]}_{fieldV~12} = - 6 E_{mech} + \frac{\mu_0}{96 \pi c^4}  \frac{d^4 I_1 (t)}{dt^4}    I_2  \oint \oint (d \vec{l_1} \cdot d \vec{l_2})   R^3_{12}.
\label{EVT}
\enq
and:
\beq
 E^{[4]}_{fieldS~12} =  E^{[4]}_{MfieldS~12c}
\label{EST}
\enq
It can easily seen that the change in volume energy is balance by the mechanical work done (see \ern{pow47}). However, the surface term remains unbalanced and cannot be balanced with a Poynting flux which indicate radiation. We stress that this term has nothing to do with the relativistic engine effect and will vanish for $\frac{d^4 I_1 (t)}{dt^4}=0$.

\subsubsection{Poynting vector}

We shall now study the Poynting vector correction of the fourth order in $\frac{1}{c}$:
\ber
\vec S'^{[4]}_{p~12} &=& \vec E'^{[0]}_1 \times \vec B'^{[4]}_2 + \vec E'^{[2]}_1 \times \vec B'^{[2]}_2 + \vec E'^{[4]}_1 \times \vec B'^{[0]}_2 + \vec E'^{[0]}_2 \times \vec B'^{[4]}_1
\nonumber \\
&+& \vec E'^{[2]}_2 \times \vec B'^{[2]}_1 +\vec E'^{[4]}_2 \times \vec B'^{[0]}_1
\label{Poyntingv4}
\enr
Taking into account that $\vec E_2^{[0]}$ is null according to \ern{E02} and $ \vec B'^{[2]}_2 $  is null according to \ern{Bsec2b}
this simplifies to:
\beq
\vec S'^{[4]}_{p~12} =  \vec E'^{[0]}_1 \times \vec B'^{[4]}_2 +  \vec E'^{[4]}_1 \times \vec B'^{[0]}_2 +
 \vec E'^{[2]}_2 \times \vec B'^{[2]}_1 +\vec E'^{[4]}_2 \times \vec B'^{[0]}_1
\label{Poyntingv4b}
\enq
We will find it useful to make the following definitions:
\ber
\vec S'^{[4]}_{p~12~a} &\equiv&  \vec E'^{[0]}_1 \times \vec B'^{[4]}_2, \quad  \vec S'^{[4]}_{p~12~b} \equiv \vec E'^{[4]}_1 \times \vec B'^{[0]}_2 ,
\nonumber \\
 \vec S'^{[4]}_{p~12~c} &\equiv& \vec E'^{[2]}_2 \times \vec B'^{[2]}_1 , \quad  \vec S'^{[4]}_{p~12~d} \equiv \vec E'^{[4]}_2 \times \vec B'^{[0]}_1.
\label{Poyntingv4c}
\enr
And thus:
\beq
\vec S'^{[4]}_{p~12} =  \vec S'^{[4]}_{p~12~a} + \vec S'^{[4]}_{p~12~b} + \vec S'^{[4]}_{p~12~c} + \vec S'^{[4]}_{p~12~d}.
\label{Poyntingv4d}
\enq
We shall also find it useful to define the Poynting flux:
\beq
PF' \equiv \oint_S \vec S'^{[4]}_{p~12} \cdot \hat n da.
\label{poyf41}
\enq
The Poynting flux will be calculated on an "infinite" sphere, while recalling that the sphere cannot actually be infinite as we are limited
by the convergence radius $R_{max}$ and can only be "big", hence:
\beq
PF' = \lim_{r \rightarrow \infty} \int \vec S'^{[4]}_{p~12} \cdot \hat r \  r^2 d \Omega.
\label{poyf42}
\enq
We will find it convenient to make the following definitions:
\ber
PF'_a  &\equiv& \oint_S \vec S'^{[4]}_{p~12~a} \cdot \hat n da, \quad
PF'_b  \equiv \oint_S \vec S'^{[4]}_{p~12~b} \cdot \hat n da, \quad
\nonumber \\
PF'_c  &\equiv& \oint_S \vec S'^{[4]}_{p~12~c} \cdot \hat n da, \quad
PF'_d  \equiv \oint_S \vec S'^{[4]}_{p~12~d} \cdot \hat n da, \quad
\label{poyf43}
\enr
And thus:
\beq
PF'  = PF'_a  + PF'_b + PF'_c  + PF'_d.
\label{poyf44}
\enq
From \ern{poyf42} it is clear that only the asymptotic expressions of  $\vec S'^{[4]}_{p~12}$ are relevant for the
Poynting flux.

Let us start by looking $\vec S'^{[4]}_{p~12~a} = \vec E'^{[0]}_1 \times \vec B'^{[4]}_2 $. The asymptotic form of
$\vec E'^{[0]}_1$ is given in \ern{E0b2}. From \ern{Bsec4b} and \ern{Efielddiv05b4v1} we have:
\beq
\vec B'^{[4]}_2 (\vec x,t) = -\frac{I_2}{2 c^2}  \oint \vec \nabla (\hat R_2 \cdot \frac{d \vec v}{dt}) \times d \tilde{\vec{l_2}}.
= -\frac{I_2}{2 c^2}  \oint \frac{1}{R_2} (\frac{d \vec v}{dt} - (\frac{d \vec v}{dt} \cdot \hat R_2) \hat R_2) \times d \tilde{\vec{l_2}}
\label{Bsec4basy}
\enq
which shall show in appendix \ref{B24asy} that for large $r$:
\beq
\vec B'^{[4]}_2 (\vec x,t) \propto \frac{1}{r^2}
\label{Bsec4basy2}
\enq
hence $\vec S'^{[4]}_{p~12~a} \propto \frac{1}{r^4}$ and thus:
\beq
 PF'_a = 0
\label{pfa}
\enq
on the infinite sphere (see \ern{poyf42}).

Let us now look at $\vec S'^{[4]}_{p~12~b} = \vec E'^{[4]}_1 \times \vec B'^{[0]}_2$
the asymptotic expression for $\vec B'^{[0]}_2$ is given in  \ern{B0b} according to which asymptotically  $\vec B'^{[0]}_2 \propto \frac{1}{r^3}$. For the electric field we turn our attention to \ern{ET4} and partition the field into relativistic engine terms and mutual inductance
(see also \ern{emui})
correction terms:
\ber
\vec E'^{[4]} &=& \vec E'^{[4]}_{rel} + \vec E'^{[4]}_{mui}
\nonumber \\
\vec E'^{[4]}_{rel} &\equiv&
\frac{3}{2 c^2} \frac{d^2 I(t)}{dt^2} \oint d \tilde{\vec{l}} \hat R \cdot \vec v
+ \frac{3}{2 c^2}  \frac{d I(t)}{dt} \oint d \tilde{\vec{l}} \hat R \cdot \frac{d \vec v}{dt}
+  \frac{I(t)}{2 c^2} \oint d \tilde{\vec{l}} \hat R \cdot \frac{d^2 \vec v}{dt^2} .
\nonumber \\
\vec E'^{[4]}_{mui} &\equiv&  -\frac{1}{24  c^4}  \frac{d^5 I (t)}{dt^5} \oint d \tilde{\vec{l}}R^3(t)
 \label{E4relmui}
\enr
This in turn will lead to a partition of $\vec S'^{[4]}_{p~12~b}$ such that:
\ber
\vec S'^{[4]}_{p~12~b} &=& \vec S'^{[4]}_{p~12~b~rel} + \vec S'^{[4]}_{p~12~b~mui}
\nonumber \\
\vec S'^{[4]}_{p~12~b~rel} &\equiv& \vec E'^{[4]}_{1~rel} \times \vec B'^{[0]}_2.
\nonumber \\
\vec S'^{[4]}_{p~12~b~mui} &\equiv& \vec E'^{[4]}_{1~mui} \times \vec B'^{[0]}_2.
\label{S4relmui}
\enr
Now $\vec E'^{[4]}_{1~rel}$ contain integrals of the type $\oint d \vec l \hat R \cdot \vec w$ for a constant $\vec w$. It follows
from \ern{hatRao} that asymptotically $\vec E'^{[4]}_{rel} \propto \frac{1}{r}$ and thus $\vec S'^{[4]}_{p~12~b~rel} \propto \frac{1}{r^4}$ thus
this term will have a null contribution to $PF_b$. Hence:
\beq
PF'_b = \lim_{r \rightarrow \infty} \int \vec S'^{[4]}_{p~12~b~mui} \cdot \hat r \  r^2 d \Omega =
\lim_{r \rightarrow \infty} \int \vec E'^{[4]}_{1~mui} \times \vec B'^{[0]}_2 \cdot \hat r \  r^2 d \Omega
\label{poyf4mui}
\enq
Plugging into the above equation, \ern{E4relmui} and \ern{B0} will result in:
\beq
PF'_b = - \frac{I_2}{24 c^4} \frac{d^5 I_1 (t)}{d t^5} \lim_{r \rightarrow \infty} \int d \Omega   \  r^2 \hat r \cdot
\oint \oint \left( d \vec l_1 \times \left( d \vec l_2 \times \frac{R_1^3 \vec R_2 }{R_2^3} \right) \right)
\label{poyf4muib}
\enq
We shall now use the definition
\beq
 \vec G_2  \equiv \atRo \vec G
\label{G2b}
\enq
 to write the above equation as:
\beq
PF'_b = - \frac{I_2}{24 c^4} \frac{d^5 I_1 (t)}{d t^5} \lim_{r \rightarrow \infty} \int d \Omega   \  r^3 \hat r \cdot
\oint \oint \left( d \vec l_1 \times \left( d \vec l_2 \times \vec G_2 \right) \right)
\label{poyf4muic}
\enq
Using standard vector identities:
\ber
 & & PF'_b =  \frac{I_2}{24 c^4} \frac{d^5 I_1 (t)}{d t^5}
\nonumber \\
 & & \lim_{r \rightarrow \infty} \int d \Omega   \  r^3
\oint \oint \left( (d \vec l_1 \cdot  d \vec l_2) (\hat r \cdot \vec G_2)  -  (\hat r \cdot  d \vec l_2) (d \vec l_1  \cdot \vec G_2)\right)
\label{poyf4muid}
\enr
This equation is analyzed in appendix \ref{PFb}. Using \ern{int13aG5} and \ern{PFbsecparj} we obtain the result:
\ber
& &   PF'_b =  - \frac{\pi}{30 c^4} \frac{d^5 I_1 (t)}{d t^5} I_2 R_{max}
\nonumber \\
& & \oint \oint \left( 7 (d \vec l_1 \cdot  d \vec l_2) (\vec x_1  \cdot \vec x_2)  + 2  (d \vec l_1  \cdot \vec x_2) (d \vec l_2  \cdot \vec x_1)\right)
\label{poyf4muie}
\enr
Turning next our attention to $PF'_c$ we notice that this term involves a cross product of $\vec E'^{[2]}_2$ and $\vec B'^{[2]}_1$.
According to \ern{E2bas2} $\vec E'^{[2]}_2$ decreases as $\frac{1}{r^3}$ while $\vec B'^{[2]}_1$ is given by \ern{Bsec2}:
\beq
\vec B'^{[2]}_1 (\vec x,t) = - \frac{1}{2 c^2}  \frac{d^2 I_1 (t)}{dt^2} \oint d \tilde{\vec{l_1}} \times \hat R_1 (t).
\label{Bsec217}
\enq
Taking into account \ern{hatRa} it then follows that $\vec B'^{[2]}_1 $ decreases asy\-mptotically as  $\frac{1}{r}$
and thus  $\vec S'^{[4]}_{p~12~c}$ decreases asymptotically as  $\frac{1}{r^4}$. It follows that:
\beq
 PF'_c = 0
\label{pfc}
\enq
Finally we evaluate $PF'_d$. We notice that this term involves a cross product of $\vec E'^{[4]}_2$ and $\vec B'^{[0]}_1$.
As for $\vec B'^{[0]}_1$, we have already indicated that  according to \ern{B0b} it behaves asymptotically as: $\vec B'^{[0]}_1 \propto \frac{1}{r^3}$. $\vec E'^{[4]}_2$ is defined in \ern{Eb4b}, it is of the form type $\oint d \vec l \hat R \cdot \vec w$ for a spatial constant $\vec w$. It thus follows from \ern{hatRao} that asymptotically $\vec E'^{[4]}_2 \propto \frac{1}{r}$ and thus $\vec S'^{[4]}_{p~12~d} \propto \frac{1}{r^4}$. We conclude that:
\beq
 PF'_d = 0
\label{pfd}
\enq
Collecting all terms of Poynting flux it follows that:
\ber
PF'  &=& PF'_a  + PF'_b + PF'_c  + PF'_d  = PF'_b =
\nonumber \\
& & \hspace{-2.3cm} - \frac{\pi}{30 c^4} \frac{d^5 I_1 (t)}{d t^5} I_2 R_{max}
 \oint \oint \left( 7 (d \vec l_1 \cdot  d \vec l_2) (\vec x_1  \cdot \vec x_2)  + 2  (d \vec l_1  \cdot \vec x_2) (d \vec l_2  \cdot \vec x_1)\right)
\label{poyf44b}
\enr
Or we may write:
\ber
& &\oint_S \vec S^{[4]}_{p~12} \cdot \hat n da = \frac{\mu_0}{(4 \pi)^2}  \oint_S \vec S'^{[4]}_{p~12} \cdot \hat n da=
- \frac{\mu_0}{480 \pi c^4} \frac{d^5 I_1 (t)}{d t^5} I_2 R_{max}
\nonumber \\
& & \oint \oint \left( 7 (d \vec l_1 \cdot  d \vec l_2) (\vec x_1  \cdot \vec x_2)  + 2  (d \vec l_1  \cdot \vec x_2) (d \vec l_2  \cdot \vec x_1)\right) .
\label{poyfl45}
\enr
This terms is clearly not a relativistic engine terms and involves a fifth derivative of the current (only a second derivative is needed for a relativistic engine). We underline again that the expansion is valid for large but finite range $ R_{max}$ defined in \ern{Rmax}, however, the explicit value of $ R_{max}$ enters explicitly only into the field energy surface terms and the Poynting flux of the fourth order in $\frac{1}{c}$, for small or null fourth derivative of the current one need not worry about such terms.

\subsubsection{Intermediate account}

We conclude that the energy \ern{Econt0ord4} of the fourth order is indeed balanced. Mechanical work invested or extracted in the system results in increase or decrease in the field energy accordingly. In the fourth order both electric and  magnetic field energies are affected by the mechanical work. The power related to the mechanical work is according to \ern{pow47}:
\beq
Power_{12}^{[4]}= - \frac{\mu_0}{96 \pi  c^2}  \frac{d^5 I_1 (t)}{dt^5} I_2  \oint  \oint (d\vec{l_1} \cdot d \vec{l_2}) R_{21}^3 +6 \frac{d E_{mech}}{dt}.
\label{pow47b}
\enq
it contains both work done by the mutual inductance and on the relativistic engine. This is equal to minus the derivative of the volume field energy \ern{EVT}:
\beq
E^{[4]}_{fieldV~12} = - 6 E_{mech} + \frac{\mu_0}{96 \pi c^4}  \frac{d^4 I_1 (t)}{dt^4}    I_2  \oint \oint (d \vec{l_1} \cdot d \vec{l_2})   R^3_{12}.
\label{EVTb}
\enq
Moreover, for the fourth order in $\frac{1}{c}$ there is also a surface contribution to the field energy given in \ern{EST}
which according to \ern{Efielddi422b} satisfies:
\beq
\frac{d E'^{[4]}_{fieldS~12}}{d t} = - \int da \ \hat n \cdot (\vec E'^{[4]}_{1~mui} \times \vec B'^{[0]}_2)
= - \oint_S \vec S^{[4]}_{p~12} \cdot \hat n da
\label{Efielddi422bb}
\enq
such that the total field energy is:
\beq
E^{[4]}_{field~12}= E^{[4]}_{fieldV~12} + E^{[4]}_{fieldS~12}
\label{Efielddiv419Tb}
\enq
The change in the field energy through the surface terms results in radiation as described by the Poynting flux depicted in \ern{poyfl45}:
\ber
& &\oint_S \vec S^{[4]}_{p~12} \cdot \hat n da = \frac{\mu_0}{(4 \pi)^2}  \oint_S \vec S'^{[4]}_{p~12} \cdot \hat n da=
- \frac{\mu_0}{480 \pi c^4} \frac{d^5 I_1 (t)}{d t^5} I_2 R_{max}
\nonumber \\
& & \oint \oint \left( 7 (d \vec l_1 \cdot  d \vec l_2) (\vec x_1  \cdot \vec x_2)  + 2  (d \vec l_1  \cdot \vec x_2) (d \vec l_2  \cdot \vec x_1)\right) .
\label{poyfl45b}
\enr
We underline that fourth order contributions are the only one that are related to the relativistic engine effect as some of the terms depends on the engine velocity $\vec v$.  If the engine is infinitely massive and no motion occurs, we are left with the mutual inductance correction terms and radiation terms which involve higher order derivatives.
For a phasor current of frequency $\omega$ \ern{CI4} indicates a relativistic correction to the classical mutual inductance which is important for large systems with high frequency.
\beq
M^{[4]}_{12} \equiv  \frac{\mu_0 \omega^4}{96 \pi c^4} \oint d \vec l_1 \cdot \oint d \vec l_2 R^3_{12}
\label{CI4b}
\enq

\section{Conclusion}

A relativistic engine is not a "perpetuum mobile" it requires energy to operate. The energy needed for its operation comes at the expanse of the electromagnetic field energy. Moreover, we have shown that the total energy required is six times the mechanical energy obtained by the engine as energy must be invested also in driving the needed current for its operation through the loops. Two times comes at the expense of the electric field energy and four times at the expense of the magnetic field energy. Notice that we have not taken resistive losses into account but if the coils are not superconductive this should be taken into account as well. As we collected all the terms up to and including
$\frac{1}{c^4}$ we have encountered for most of the time terms that can be thought of as relativistic corrections to the mutual inductance formula and are not connected in any way to the relativistic engine effect. For order $\frac{1}{c^3}$ there are also radiation losses which may be avoided by cleverly constructing the loop coils orthogonal to each other. For higher $\frac{1}{c^4}$  the nature of our series expansion prevents us from evaluating the radiation flux at infinity and we must suffice with the radiation flux over a sphere of radius $R_{max}$ which is the distance after which our approximation becomes invalid.

In this work we have only dealt with the energy exchange due to the interaction of two loops but of course even a single loop looses energy due to radiation. Future works will consider other relativistic concepts such as an electric (rather than magnetic) relativistic engine. We will also be interested in studying the relativistic ramifications of small body moving in a large structure generating an "external field". In such a body
we expect additional contribution to the main "classical" force which are due to the relativistic retardation.

Finally we remark that although an energy of $6 E_{mech}$ seems excessive and inefficient, it is highly efficient with respect to other types of engines which are purely electromagnetic. For example to reach a momentum $p$ using a photon engine one needs an energy of $E_p=pc$ while for a relativistic engine an energy of $E_r = 3 p v  $ will suffice. The ratio is $\frac{E_p}{E_r } = \frac{c}{3 v}$ which is a huge number for non relativistic speeds.

\vspace{0.5cm}
\noindent
{\LARGE \bf Acknowledgement}
\vspace{0.5cm}

\noindent This ‎research was supported by the U.S. Department of Energy (DE-AC02-09CH11466).
\\
\\
\noindent Shailendra Rajput is thankful to the Israeli Council for Higher Education (CHE) for
fellowship.

\vfill \eject
\vspace{0.5cm}
\noindent ‎
{\LARGE \bf Appendix}

\appendix

\section{On the nullification of a certain integral}

We would like to prove the equality:
\beq
\int d^3 x  \oint \oint [ (d \vec l_1 \cdot \vec \nabla \frac{1}{R_2 (t)}) (d \vec l_2  \cdot \vec \nabla \frac{1}{R_1 (t)}) ]
= 0
\label{Efielddiv05b}
\enq
Which may be written in terms of the Einstein summation convention as follows:
\beq
\int d^3 x  \oint \oint [ (d  l_{1m} \partial_m \frac{1}{R_2 (t)}) (d  l_{2n}  \partial_n \frac{1}{R_1 (t)}) ]
= 0
\label{Efielddiv05c}
\enq
in the above we have used the symbol  $\partial_m = \frac{\partial}{\partial x_m}$.
To do this let us look at a more general case.  Let $g (\vec R_1)$ and $h (\vec R_2)$ be two arbitrary functions.
And let us evaluate the integral:
\beq
Int =\int d^3 x  \oint \oint  (d  l_{1m} \partial_m h (\vec R_2)) (d  l_{2n}  \partial_n g (\vec R_1))
\label{Efielddiv05d}
\enq
 The following set of equations follow:
\ber
 & & \partial_m g (\vec R_1) \partial_n h (\vec R_2)  = \partial_n ( \partial_m g (\vec R_1)  h (\vec R_2) )
  - h (\vec R_2) \partial^2_{mn} g (\vec R_1)
  \nonumber \\
  &=& \partial_n ( \partial_m g (\vec R_1)  h (\vec R_2) ) - \partial_m ( \partial_n g (\vec R_1)  h (\vec R_2) )
  +\partial_m h (\vec R_2) \partial_n g (\vec R_1)
 \enr
Now since $\vec R_1 = \vec x - \vec x_1$ it follows that:
\beq
\partial_n g (\vec R_1) = -\partial_{n1} g (\vec R_1), \qquad \partial_{n1} \equiv \frac{\partial}{\partial x_{1n}}
\label{parjg}
\enq
And since $\vec R_2 = \vec x - \vec x_2$ it follows that:
\beq
\partial_m h (\vec R_2) = -\partial_{m2} h (\vec R_2), \qquad \partial_{m2} \equiv \frac{\partial}{\partial x_{2m}}
\enq
Summing up the above results we have:
\ber
 & & \partial_m g (\vec R_1) \partial_n h (\vec R_2) =
  \nonumber \\
  & & \partial_n ( \partial_m g (\vec R_1)  h (\vec R_2) ) - \partial_m ( \partial_n g (\vec R_1)  h (\vec R_2) )
  +\partial_{m2} h (\vec R_2) \partial_{n1} g (\vec R_1)
 \enr
We thus conclude that:
\ber
Int &=&\int d^3 x  \oint \oint d  l_{1n} d  l_{2m} [\partial_n ( \partial_m g (\vec R_1)  h (\vec R_2) ) - \partial_m ( \partial_n g (\vec R_1)  h (\vec R_2) )
 \nonumber \\
&+& \partial_{m2} h (\vec R_2) \partial_{n1} g (\vec R_1)]
\label{Efielddiv05e}
\enr
Now for every single valued set of function $h$ and $g$ we have:
\beq
\oint d  l_{1n} \partial_{n1} g (\vec R_1) =  \oint d  g  = 0, \quad
\oint d  l_{2m} \partial_{m2} h (\vec R_2) =  \oint d  h  = 0
\label{oindl}
\enq
Hence:
\beq
Int = \int d^3 x  \oint \oint d  l_{1n} d  l_{2m} [\partial_n ( \partial_m g (\vec R_1)  h (\vec R_2) ) - \partial_m ( \partial_n g (\vec R_1)  h (\vec R_2) )]
\label{Efielddiv05f}
\enq
The above integral contains only gradients which once integrated can only contribute to surface terms as follows:
\beq
Int =  \oint \oint d  l_{1n} d  l_{2m} [\oint da_n  \partial_m g (\vec R_1)  h (\vec R_2)  - \oint da_m \partial_n g (\vec R_1)  h (\vec R_2)]
\label{Efielddiv05g}
\enq
Now:
\ber
& & \oint \oint d  l_{1n} d  l_{2m} \oint da_m  \partial_n g (\vec R_1)  h (\vec R_2)
 \nonumber \\
 &=& -\oint \oint d  l_{1n} d  l_{2m} \oint da_m  \partial_{n1} g (\vec R_1)  h (\vec R_2) = 0
\label{Efielddiv05gb}
\enr
according to \ern{parjg} and \ern{oindl}. Hence:
\beq
Int =  \oint \oint d  l_{1n} d  l_{2m} \oint da_n  \partial_m g (\vec R_1)  h (\vec R_2)
\label{Efielddiv05gc}
\enq
Suppose now that the system is contained in an infinite sphere of radius $r=\infty$ and suppose that:
\beq
\lim_{r \rightarrow \infty} \oint da_n  \partial_m g (\vec R_1)  h (\vec R_2)   = 0
\label{Efielddiv05gd}
\enq
Then it follows that:
\beq
Int =  0
\label{Efielddiv05h}
\enq
And \ern{Efielddiv05b} is proved.
To verify that this is indeed so we only need to substitute  $h (\vec R_2) = \frac{1}{R_2 (t)}$ and
 $g(\vec R_1) = \frac{1}{R_1 (t)}$ and take into account \ern{spherele}, \ern{asym3} and \ern{asym5}.

\section{Asymptotic values}
\label{asymval}

In the limit  $r \rightarrow \infty$ we may think of:
\beq
\epsilon \equiv \frac{1}{r}
\label{eps}
\enq
 as a small parameter. In
terms of this small parameter we expand $\vec G$ defined in \ern{Gdef} :
\beq
\vec G(\epsilon) =\vec G(0) + \epsilon \vec G'(0) + \frac{1}{2} \epsilon^2 \vec G''(0) + O \left( \epsilon^3 \right)
\label{Gexp}
\enq
Inserting \ern{Gexp} into \ern{Efielddiv311c} will result in:
\ber
& & \oint \oint (d \tilde{\vec{l_1}} \cdot d \tilde{\vec{l_2}}) \lim_{r \rightarrow \infty} \int d \Omega \ r^2 \hat r \cdot \vec G
=\int d \Omega  \lim_{\epsilon \rightarrow 0} \frac{1}{\epsilon^2} \oint \oint d \vec{l_1} \cdot d \vec{l_2} \hat r \cdot \vec G (0)
\nonumber \\
&+ & \int d \Omega  \lim_{\epsilon \rightarrow 0} \frac{1}{\epsilon} \oint \oint d \vec{l_1} \cdot d \vec{l_2} \hat r \cdot \vec G' (0)
+ \int d \Omega  \frac{1}{2} \oint \oint d \vec{l_1} \cdot d \vec{l_2} \hat r \cdot \vec G'' (0)
\label{Efielddiv311c1}
\enr
the terms $O \left( \epsilon^3 \right)$ will cancel in the limit. Obvious the first two terms in right hand side will diverge unless the closed loop integrals vanish.
Let us denote:
\beq
\tR \equiv \frac{\vec R}{r} = \hat r -\epsilon \vec x'
\label{tR}
\enq
and
\beq
\atR = \sqrt{1-2 \epsilon \hat r \cdot \vec x' + \epsilon^2 \vec x'^2}
\label{atR}
\enq
in terms of $\tR$ we may write $\vec G$ as:
\beq
\vec G  =  \frac{ R_1^2 \vec R_2}{R_2^3 } = \frac{ \atRo^2 \tR_2}{\atRt^3 }
\label{Gdef2}
\enq
obviously:
\beq
\vec G (0) = \hat r.
\label{G1}
\enq
However, since a closed loop integral over a constant vanishes  (see \ern{lopC}), it follows that:
\beq
\int d \Omega  \lim_{\epsilon \rightarrow 0} \frac{1}{\epsilon^2} \oint \oint d \vec{l_1} \cdot d \vec{l_2} \hat r \cdot \vec G (0)
= \int d \Omega  \lim_{\epsilon \rightarrow 0} \frac{1}{\epsilon^2} \oint \oint d \vec{l_1} \cdot d \vec{l_2} = 0
\label{intG1}
\enq
Now let us calculate $G'(\epsilon)$  to do this we notice the following identities:
\beq
\frac{d \tR}{d \epsilon} = - \vec x', \qquad \frac{d^2 \tR}{d \epsilon^2} = 0, \qquad
\frac{d \atR}{d \epsilon} = \frac{\epsilon \vec x'^2 - \hat r \cdot \vec x'}{\atR}
\label{tilident}
\enq
Using the above identities we calculate:
\beq
\vec G'(\epsilon) = 2 (\epsilon \vec x_1^2 - \hat r \cdot \vec x_1) \tR_2 \atRt^{-3}
 - 3 (\epsilon \vec x_2^2 - \hat r \cdot \vec x_2)  \tR_2 \atRt^{-5} \atRo^{2}
 - \vec x_2 \atRt^{-3} \atRo^{2}
\label{Gder}
\enq
And thus:
\beq
\vec G'(0) = \hat r \left(3 (\hat r \cdot \vec x_2) - 2 (\hat r \cdot \vec x_1)\right) - \vec x_2
\label{Gder0}
\enq
Hence:
\ber
& &\int d \Omega  \lim_{\epsilon \rightarrow 0} \frac{1}{\epsilon} \oint \oint d \vec{l_1} \cdot d \vec{l_2} \hat r \cdot \vec G' (0)
\nonumber \\
&=& 2 \int d \Omega  \lim_{\epsilon \rightarrow 0} \frac{1}{\epsilon} \oint \oint d \vec{l_1} \cdot d \vec{l_2}
\left((\hat r \cdot \vec x_2) -  (\hat r \cdot \vec x_1)\right) = 0
\label{intG2}
\enr
because at least one of the two loop integrals is done over a constant. Finally we calculate $\vec G''(\epsilon)$ which leads to a somewhat lengthy but straight forward expression:
\ber
\vec G''(\epsilon) &=& 2 \vec x_1^2 \tR_2 \atRt^{-3}
\nonumber \\
&+& 2 (\epsilon \vec x_1^2 - \hat r \cdot \vec x_1)
\left( - \vec x_2  \atRt^{-3} - 3 \tR_2 \atRt^{-5} (\epsilon \vec x_2^2 - \hat r \cdot \vec x_2) \right)
\nonumber \\
&+& 15 \atRt^{-7} (\epsilon \vec x_2^2 - \hat r \cdot \vec x_2)^2 \atRo^{2} \tR_2
\nonumber \\
&-& 6 \atRt^{-5} (\epsilon \vec x_1^2 - \hat r \cdot \vec x_1)  \tR_2 (\epsilon \vec x_2^2 - \hat r \cdot \vec x_2)
\nonumber \\
&+& 3 \vec x_2 \atRt^{-5} \atRo^{2}   (\epsilon \vec x_2^2 - \hat r \cdot \vec x_2)
\nonumber \\
&-& 3 \atRt^{-5} \atRo^{2} \vec x_2^2 \tR_2
\nonumber \\
&-& \vec x_2 \left(2 (\epsilon \vec x_1^2 - \hat r \cdot \vec x_1) \atRt^{-3}  - 3 \atRo^{2} \atRt^{-5} (\epsilon \vec x_2^2 - \hat r \cdot \vec x_2)\right)
\label{Gsder}
\enr
From the above expression we calculate $\vec G''(0)$ as follows:
\beq
\vec G''(0) = 4 (\hat r \cdot \vec x_1)\vec x_2 + 2 \vec x_1^2 \hat r + 15(\hat r \cdot \vec x_2)^2 \hat r
 -6 \vec x_2 (\hat r \cdot \vec x_2) - 3 \vec x_2^2 \hat r
\label{Gsder0}
\enq
Obviously only the first term contributes as it depends on both $\vec x_1$ and $\vec x_2$ (and not on each variable alone) and hence:
\beq
 \int d \Omega  \frac{1}{2} \oint \oint d \vec{l_1} \cdot d \vec{l_2} \hat r \cdot \vec G'' (0)
 = 2  \int d \Omega  \oint \oint d \vec{l_1} \cdot d \vec{l_2} (\hat r \cdot \vec x_2) (\hat r \cdot \vec x_1)
\label{Efielddiv311c2}
\enq
Finally inserting the results from \ern{intG1}, \ern{intG2} and \ern{Efielddiv311c2} into \ern{Efielddiv311c1} we obtain
\beq
\oint \oint (d \tilde{\vec{l_1}} \cdot d \tilde{\vec{l_2}}) \lim_{r \rightarrow \infty} \int d \Omega \ r^2 \hat r \cdot \vec G
=2 \int d \Omega   \oint \oint d \vec{l_1} \cdot d \vec{l_2} (\hat r \cdot \vec x_2) (\hat r \cdot \vec x_1)
\label{Efielddiv311c3}
\enq
which is identical to \ern{Efielddiv311c}.

\section{Q function evaluation}
\label{Qfun}

We shall now calculate the term:
\beq
\vec Q = \int \left[\frac{1}{R_1(t)} \frac{\vec R_2(t)}{R_2^3(t)} \right] d^3 x
\label{Gt}
\enq
First let us introduce a change of variables:
\beq
\vec y = \vec R_2 = \vec x - \vec x_2
\label{y}
\enq
Since the integral $\vec Q$ is calculated at a fixed point $\vec x_2$ it follows that $d^3 y =  d^3 x$ and:
\beq
\vec R_1 = \vec x - \vec x_1 = \vec y + \vec x_2 - \vec x_1 = \vec y - \vec R_{12}.
\label{R1}
\enq
This leads to the following expression for $\vec Q$:
\beq
\vec Q = \int y^{-3} \vec y \left|\vec y - \vec R_{12}\right|^{-1} d^3 y .
\label{G2Q}
\enq
This integral is now evaluated using a spherical coordinate system in which the "z" axis point
at the direction of $\vec R_{21}$. In this case $d^3 y = - y^2 dy d \cos \theta d \phi$ and $\vec Q$
can be calculated as follows:
\beq
\vec Q = -\int_{0}^{\infty}dy \int_{1}^{-1} d \cos \theta \int_{0}^{2 \pi} d \phi
 y^{-1} \vec y \left|\vec y - \vec R_{12}\right|^{-1} .
\label{G3}
\enq
Now:
\beq
\left|\vec y - \vec R_{12}\right| = \sqrt{y^2 + R_{12}^2 - 2 \vec y \cdot \vec R_{12}} = \sqrt{y^2 + R_{12}^2 - 2  y  R_{12} \cos \theta},
\label{yR}
\enq
In which we notice that the above expression is not dependent on the azimuthal angel $\phi$. Moreover, using
a cartesian set of unit vectors $\hat{y_1},\hat{y_2},\hat{y_3}$ one may write:
\beq
y^{-1} \vec y = \sin \theta \cos \phi \hat{y_1} +  \sin \theta \sin \phi \hat{y_2} + \cos \theta \hat{y_3},
\label{unity}
\enq
Thus it can easily be seen that there is component to $\vec Q$ in the $\hat{y_1},\hat{y_2}$ directions as the
azimuthal integral vanishes. In the $\hat{y_3}$ direction the azimuthal integral is trivial and we obtain the result:
\beq
\vec Q =2 \pi \hat{y_3} \int_{0}^{\infty}dy \int_{-1}^{1} d \cos \theta  \cos \theta
 \sqrt{y^2 + R_{12}^2 - 2  y  R_{12} \cos \theta}^{~-1} .
\label{G4}
\enq
Let us make a change of variables $s \equiv \cos \theta, y' \equiv \frac{y}{R_{12}}$ and notice that $\hat{y_3}= \hat{R}_{12}$ which
is a unit vector in the direction of $\vec R_{12}$, in terms of those variables we obtain a simpler representation of $\vec Q$:
\beq
\vec Q =2 \pi \hat{R}_{21} \int_{0}^{\infty}dy' \int_{-1}^{1} d s  s  \sqrt{y'^2 + 1 - 2  y' s}^{~-1} .
\label{G5}
\enq
However, we can evaluate analytically the $s$ integral to obtain:
\beq
\int_{-1}^{1} d s  s  \sqrt{y'^2 + 1 - 2  y' s}^{~-1} = \frac{2}{3} \left\{ \begin{array}{cc}
                                                                               \frac{1}{y'^2} & y' \geq 1 \\
                                                                               y' & y' < 1
                                                                             \end{array} \right. .
\label{sint}
\enq
And plugging this back into \ern{G5} we obtain:
\beq
\vec Q  = 2 \pi \hat{R}_{12}.
\label{G6}
\enq

\section{Asymptotic form of $\vec B'^{[4]}_2$ }
\label{B24asy}

Let us write \ern{hatRa} in the form (keeping only first order terms in $\frac{1}{r}$):
\beq
\hat R_{2} \simeq \hat r - \frac{\vec x_{2\bot}}{r}
\label{hatRao}
\enq
in which we define a perpendicular vector to $\hat r$ as follows:
\beq
\vec w_{\bot} =\vec w - \hat r (\hat r \cdot \vec w)
\label{wper}
\enq
Hence we may write up to first order in $\frac{1}{r}$):
\beq
(\frac{d \vec v}{dt} \cdot \hat R_2) \hat R_2 \simeq (\frac{d \vec v}{dt} \cdot \hat r) \hat r -\frac{\vec x_{2\bot}}{r} (\frac{d \vec v}{dt} \cdot \hat r) -  -\frac{\hat r}{r} (\frac{d \vec v}{dt} \cdot \vec x_{2\bot})
\label{auxcal}
\enq
Let us use the above equation and \ern{asym3} in \ern{Bsec4basy}, we obtain:
\ber
& & \oint \frac{1}{R_2} (\frac{d \vec v}{dt} - (\frac{d \vec v}{dt} \cdot \hat R_2) \hat R_2) \times d \vec{l_2}
\nonumber \\
&\simeq& \frac{1}{r} \oint (1 + \frac{\hat r \cdot \vec x_2}{r}) \left(\frac{d \vec v}{dt}_{\bot} + \frac{\vec x_{2\bot}}{r} (\frac{d \vec v}{dt} \cdot \hat r) + \frac{\hat r}{r} (\frac{d \vec v}{dt} \cdot \vec x_{2\bot})\right) \times d \vec{l_2}
\label{Bsec4basy22}
\enr
Now since for a constant loop integral we have:
\beq
\oint \vec C \times d \vec{l}  = 0
\label{conlocrin}
\enq
It follows that:
\ber
& & \oint \frac{1}{R_2} (\frac{d \vec v}{dt} - (\frac{d \vec v}{dt} \cdot \hat R_2) \hat R_2) \times d \vec{l_2}
\nonumber \\
&\simeq& \frac{1}{r^2} \oint \left( (\hat r \cdot \vec x_2)\frac{d \vec v}{dt}_{\bot} + \vec x_{2\bot}(\frac{d \vec v}{dt} \cdot \hat r) + \hat r(\frac{d \vec v}{dt} \cdot \vec x_{2\bot})\right) \times d \vec{l_2}
\label{Bsec4basy23}
\enr
and \ern{Bsec4basy2} is thus derived.

\section{Evaluating $PF_b$}
\label{PFb}

Let us look at $\vec G_2$ defined in \ern{G2b}. In terms of the small parameter $\epsilon$ defined in \ern{eps} we may expand $\vec G_2$ as follows:
\beq
\vec G_2 (\epsilon) =\vec G_2 (0) + \epsilon \vec G'_2(0) + \frac{1}{2} \epsilon^2 \vec G''_2 (0) + \frac{1}{6} \epsilon^3 \vec G'''_2 (0)
+ O \left( \epsilon^4 \right)
\label{Gexp2}
\enq
Thus we have:
\ber
& & \oint \oint (d \vec{l_1} \cdot d \vec{l_2}) \lim_{r \rightarrow \infty} \int d \Omega \ r^3 \hat r \cdot \vec G_2
\nonumber \\
&=& \oint \oint (d \vec{l_1} \cdot d \vec{l_2}) \lim_{\epsilon \rightarrow 0} \frac{1}{\epsilon^3}  \int d \Omega \hat r  \cdot \vec G_2 (0)
\nonumber \\
&+& \oint \oint (d \vec{l_1} \cdot d \vec{l_2}) \lim_{\epsilon \rightarrow 0} \frac{1}{\epsilon^2}  \int d \Omega \hat r  \cdot \vec G'_2 (0)
\nonumber \\
&+& \frac{1}{2}\oint \oint (d \vec{l_1} \cdot d \vec{l_2}) \lim_{\epsilon \rightarrow 0} \frac{1}{\epsilon}  \int d \Omega \hat r  \cdot \vec G''_2 (0)
\nonumber \\
&+& \frac{1}{6}\oint \oint (d \vec{l_1} \cdot d \vec{l_2}) \int d \Omega \hat r  \cdot \vec G'''_2 (0)
\label{int13a1}
\enr
the terms $O \left( \epsilon^4 \right)$ will cancel in the limit. Obviously the first three terms in right hand side will diverge unless the closed loop integrals vanish. However:
\beq
\vec G_2 (0) = \hat r.
\label{G2a1}
\enq
Since a closed loop integral over a constant vanishes  (see \ern{lopC}), it follows that:
\beq
\int d \Omega  \lim_{\epsilon \rightarrow 0} \frac{1}{\epsilon^3} \oint \oint (d \vec{l_1} \cdot d \vec{l_2}) \hat r \cdot \vec G_2 (0)
= \int d \Omega  \lim_{\epsilon \rightarrow 0} \frac{1}{\epsilon^3} \oint \oint (d \vec{l_1} \cdot d \vec{l_2})= 0
\label{int13a2}
\enq
For calculating the derivatives of $\vec G_2$ we shall use computer algebra due to the complexity of the expressions. We
shall write down only the values of the derivative in $0$ since only those are of interest to us for evaluating the expressions in
\ern{int13a1}. For $\vec G'_2 (0) $ we obtain:
\beq
\vec G'_2 (0) = 3 \hat r \left(\hat r \cdot (\vec x_2 - \vec x_1) \right) - \vec x_2
\label{G2der0}
\enq
Hence:
\ber
& &\int d \Omega  \lim_{\epsilon \rightarrow 0} \frac{1}{\epsilon^2} \oint \oint d \vec{l_1} \cdot d \vec{l_2} \hat r \cdot \vec G' (0)
\nonumber \\
&=&  \int d \Omega  \lim_{\epsilon \rightarrow 0} \frac{1}{\epsilon^2} \oint \oint d \vec{l_1} \cdot d \vec{l_2}
\left(2(\hat r \cdot \vec x_2) -  3(\hat r \cdot \vec x_1)\right) = 0
\label{int13aG2}
\enr
because at least one of the two loop integrals is done over a constant. Let us now evaluate $\vec G''_2 (0)$:
\ber
\vec G''_2(0) &=& 3 \left ( \hat{r} (\hat{r}\cdot \vec x_1)^2 - 2 \vec x_2 (\hat{r}\cdot \vec x_2) + 5 \hat{r} (\hat{r}\cdot \vec x_2)^2 \right.
\nonumber \\
&+& \left. 2 (\hat{r}\cdot \vec x_1) (\vec x_2 - 3 \hat{r} (\hat{r}\cdot \vec x_2)) + \hat{r} (\vec x_1^2 - \vec x_2^2)
 \right)
\label{G2sder0}
\enr
Obviously only the first term contributes as it depends on both $\vec x_1$ and $\vec x_2$ (and not on each variable alone) thus:
\beq
\vec G''_{2~12}(0) = 6  (\hat{r}\cdot \vec x_1) (\vec x_2 - 3 \hat{r} (\hat{r}\cdot \vec x_2))
\label{G2sder0b}
\enq
and hence:
\ber
&&\frac{1}{2}\oint \oint (d \vec{l_1} \cdot d \vec{l_2}) \lim_{\epsilon \rightarrow 0} \frac{1}{\epsilon}  \int d \Omega \hat r  \cdot \vec G''_2 (0)
\nonumber \\
&=& -6 \lim_{\epsilon \rightarrow 0} \frac{1}{\epsilon}   \oint \oint (d \vec{l_1} \cdot d \vec{l_2} ) \int d \Omega
(\hat r \cdot \vec x_2) (\hat r \cdot \vec x_1)
\label{int13aG3}
\enr
Taking into account \ern{rnrk} we thus have:
\beq
\frac{1}{2}\oint \oint (d \vec{l_1} \cdot d \vec{l_2}) \lim_{\epsilon \rightarrow 0} \frac{1}{\epsilon}  \int d \Omega \hat r  \cdot \vec G''_2 (0)
= -8 \pi \lim_{\epsilon \rightarrow 0} \frac{1}{\epsilon}   \oint \oint (d \vec{l_1} \cdot d \vec{l_2} ) \vec x_1 \cdot \vec x_2.
\label{int13aG4}
\enq
As this result is diverging we are reminded that we are not allowed to use our expansion beyond $R_{max}$ (see \ern{Rmax}) and thus
the above integral takes the value:
\beq
- 8 \pi R_{max}   \oint \oint (d \vec{l_1} \cdot d \vec{l_2} ) \vec x_1 \cdot \vec x_2 + O(1).
\label{int13aG5}
\enq
in which $ O(1)$ stands for terms which are either independent of $R_{max}$ or decrease with $R_{max}$.

For the same reasons the first two terms in the expansion do not contribute to the second integral of \ern{poyf4muid} and we are left with third term. Thus:
\beq
-  \lim_{r \rightarrow \infty} \int d \Omega   \  r^3 \oint \oint  \left( (\hat r \cdot  d \vec l_2) (d \vec l_1  \cdot \vec G_2)\right)
\label{PFbsecparc}
\enq
will be equal to:
\beq
- \frac{1}{2} \lim_{\epsilon \rightarrow 0} \frac{1}{\epsilon } \int d \Omega  \oint \oint  \left( (\hat r \cdot  d \vec l_2) (d \vec l_1  \cdot \vec G''_2 (0) )\right)
\label{PFbsecpard}
\enq
and taking into account \ern{G2sder0b} we obtain:
\beq
- 3 \lim_{\epsilon \rightarrow 0} \frac{1}{\epsilon } \int d \Omega  \oint \oint  \left( (\hat r \cdot  d \vec l_2) d \vec l_1  \cdot (\hat{r}\cdot \vec x_1) \left(\vec x_2 - 3 \hat{r} (\hat{r}\cdot \vec x_2)\right)\right)
\label{PFbsecpare}
\enq
The above expression can be written in terms of Einstein summation notation as:
\beq
- 3 \lim_{\epsilon \rightarrow 0} \frac{1}{\epsilon } \int d \Omega  \oint \oint \left[  (d \vec l_1  \cdot \vec x_2) x_{1k} dl_{2n}
\hat{r}_k \hat{r}_n - 3 dl_{1m} dl_{2n} x_{1s} x_{2k} \hat{r}_m \hat{r}_n  \hat{r}_s  \hat{r}_k \right]
\label{PFbsecparf}
\enq
taking into account \ern{rnrk} and also the result \cite{Jackson}:
 \beq
\oint d \Omega  \hat r_n \hat r_k \hat r_m \hat r_l = \frac{4 \pi}{15} (\delta_{nk} \delta_{lm} + \delta_{nl} \delta_{km}
 +\delta_{nm} \delta_{kl} )
\label{rnrkrmrl}
\enq
we obtain:
\ber
&-& 4 \pi \lim_{\epsilon \rightarrow 0} \frac{1}{\epsilon } \oint \oint \left[  (d \vec l_1  \cdot \vec x_2) x_{1k} dl_{2n}
\delta_{nk} \right.
\nonumber \\
&-& \left. \frac{3}{5} dl_{1m} dl_{2n} x_{1s} x_{2k} (\delta_{nk} \delta_{sm} + \delta_{ns} \delta_{km}
 +\delta_{nm} \delta_{ks} ) \right].
\label{PFbsecparg}
\enr
Taking into account  that $\oint d \vec l_1  \cdot \vec x_1 = \oint d \vec l_2  \cdot \vec x_2 = 0$, \ern{PFbsecparg} takes the form:
\ber
&-&4 \pi \lim_{\epsilon \rightarrow 0} \frac{1}{\epsilon } \oint \oint \left[  (d \vec l_1  \cdot \vec x_2) (d \vec l_2  \cdot \vec x_1)  \right.
\nonumber \\
&-& \left. \frac{3}{5} ((d \vec l_1  \cdot \vec x_2) (d \vec l_2  \cdot \vec x_1) + (d \vec l_1  \cdot d \vec l_2)  (\vec x_1  \cdot \vec x_2) ) \right]
\label{PFbsecparh}
\enr
or more simply as:
\beq
\frac{4 \pi }{5} \lim_{\epsilon \rightarrow 0} \frac{1}{\epsilon } \oint \oint \left[3 (d \vec l_1  \cdot d \vec l_2)  (\vec x_1  \cdot \vec x_2) -2 (d \vec l_1  \cdot \vec x_2) (d \vec l_2  \cdot \vec x_1)  \right]
\label{PFbsecpari}
\enq
This term is diverging and thus we shall replace the infinite sphere with a sphere which is simply big to the maximum allowable size of the
expansion:
\beq
\frac{4 \pi }{5} R_{max} \oint \oint \left[3 (d \vec l_1  \cdot d \vec l_2)  (\vec x_1  \cdot \vec x_2) -2 (d \vec l_1  \cdot \vec x_2) (d \vec l_2  \cdot \vec x_1)  \right] + O(1).
\label{PFbsecparj}
\enq

\end{document}